\documentclass[letterpaper,twocolumn,10pt]{article}
\usepackage{usenix2019_v3_arxiv}

\usepackage{tikz}
\usepackage{amsmath}
\usepackage{xspace}
\usepackage{amsfonts, amssymb}
\usepackage{mathtools}
\usepackage{subcaption}
\usepackage{multirow}
\usepackage{algorithm}
\usepackage{booktabs}
\usepackage[noend]{algpseudocode}
\usepackage{enumitem}
\usepackage{titlesec}
\usepackage[T1]{fontenc}
\usepackage{inconsolata}
\usepackage{makecell}
\usepackage{comment}
\usepackage{makecell}
\usepackage[most]{tcolorbox}
\usepackage{capt-of}

\usepackage{listings}
\usepackage{color}

\definecolor{dkgreen}{rgb}{0,0.6,0}
\definecolor{gray}{rgb}{0.5,0.5,0.5}
\definecolor{mauve}{rgb}{0.58,0,0.82}

\usepackage{amsthm}

\newtheoremstyle{tight}  
  {2pt}                  
  {2pt}                  
  {\itshape}             
  {}                     
  {\bfseries}            
  {}                    
  { }                    
  {}                     

\theoremstyle{tight}
\newtheorem{theorem}{Theorem}

\newcommand{\PHB}[1]{\noindent\textbf{#1}}
\newcommand{\PHM}[1]{\vspace{.4em}\noindent\textbf{#1}} 
\newcommand{\SystemName}{\textsc{RollMux}\xspace}

\begin{document}
\setlength{\floatsep}{3pt plus 2pt minus 2pt}
\setlength{\textfloatsep}{3pt plus 2pt minus 2pt}
\setlength{\intextsep}{3pt plus 2pt minus 2pt}
\setlength{\abovecaptionskip}{3pt plus 1pt minus 1pt}
\setlength{\belowcaptionskip}{3pt plus 1pt minus 1pt}
\setlength{\abovedisplayskip}{3pt}
\setlength{\belowdisplayskip}{3pt}


\date{}




\title{ \Large \bf \SystemName: Phase-Level Multiplexing for Disaggregated RL Post-Training}


\author{
  Tianyuan Wu$^{\dagger}$, 
  Lunxi Cao$^{\dagger}$,
  Yining Wei$^{\ddagger}$,
  Wei Gao$^{\dagger}$,
  Yuheng Zhao$^{\dagger}$,
  Dakai An$^{\dagger}$,
  Shaopan Xiong$^{\S}$,\\
  Zhiqiang Lv$^{\S}$,
  Ju Huang$^{\S}$,
  Siran Yang$^{\S}$,
  Yinghao Yu$^{\S}$,
  Jiamang Wang$^{\S}$,
  Lin Qu$^{\S}$, 
  Wei Wang$^{\dagger}$ \\
  $^{\dagger}$Hong Kong University of Science and Technology, $^{\ddagger}$UIUC, $^{\S}$Alibaba Group
}

\maketitle


\begin{abstract}

Rollout–training disaggregation is emerging as the standard architecture for
Reinforcement Learning (RL) post-training, where memory-bound rollout and
compute-bound training are physically disaggregated onto purpose-built clusters
to maximize hardware efficiency. However, the strict synchronization required
by on-policy algorithms introduces severe \emph{dependency bubbles}, forcing
one cluster to idle while the dependent phase is running on the other.
We present \SystemName{}, a cluster scheduling framework that reclaims these
bubbles through cross-cluster orchestration. \SystemName{} is built on the
insight that the structural idleness of one job can be effectively utilized
by the active phase of another. To realize this, we introduce the \emph
{co-execution group} abstraction, which partitions the cluster into isolated
locality domains. This abstraction enables a \emph{two-tier scheduling
architecture}: an \emph{inter-group scheduler} that optimizes job placement
using conservative stochastic planning, and an \emph{intra-group scheduler}
that orchestrates a provably optimal round-robin schedule. The
group abstraction also imposes a \emph{residency constraint}, ensuring that massive model
states remain cached in host memory to enable ``warm-start''
context switching. We evaluate \SystemName{} on a production-scale testbed
with 328 H20 and 328 H800 GPUs. \SystemName{} improves cost efficiency by 1.84$\times$ over
standard disaggregation and 1.38$\times$ over state-of-the-art
co-located baselines, all while achieving 100\% SLO attainment.

\end{abstract}


\section{Introduction}
\label{sec:intro}

The focus of Large Language Model (LLM) development has shifted from
pre-training to \emph{Reinforcement Learning (RL) post-training}~\cite
{deepseekr1, qwq32, openaio4}, a critical technique for unlocking reasoning
capabilities in complex domains such as mathematics~\cite
{maxwell_jia_2024_aime_dataset}, coding~\cite
{pan2024trainingsoftwareengineeringagents}, and tool use~\cite
{yao2022webshop, feng2025retool}. To achieve optimal performance and model
stability, production practices have converged on \emph
{synchronous, on-policy algorithms}~\cite
{shao2024deepseekmath, rastogi2025magistral, deepseekr1}. This paradigm
mandates a strict, iterative learning process comprising three phases with distinct resource bottlenecks:
(1) \textbf{rollout}, a \emph{memory-bandwidth-bound} inference stage where
the model generates tokens as experience trajectories; (2) \textbf{training}, a \emph{compute-intensive} stage where model
parameters are optimized based on the rewards; and (3) \textbf
{synchronization}, a \emph{network-bound} stage where updated model
parameters are propagated back to inference workers.

\begin{figure}[tb]
    \centering
    \includegraphics[width=0.95\linewidth]{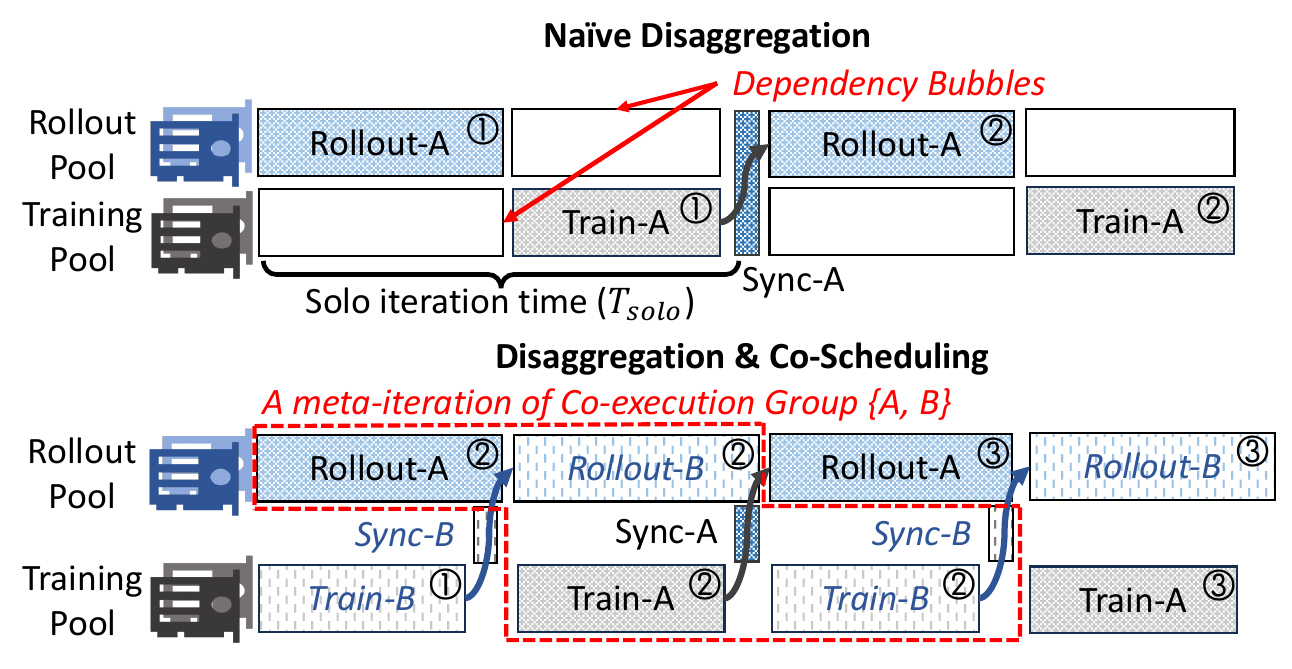}
    \caption{Comparison between existing disaggregated RL architecture and \SystemName's co-scheduling paradigm.}
    \label{fig:core_insight}
\end{figure}

To accommodate the divergent resource demands of these phases, production
deployment increasingly employs a \emph{disaggregated architecture}~\cite
{zhong2025streamrl,fu2025areal,wang2025seamlessflow,yan2025arealhex}. 
Unlike monolithic provisioning, this architecture separates the rollout and training phases onto purpose-built clusters (see \autoref{fig:core_insight}): a \emph{rollout pool}
consisting of cost-effective, inference-optimized GPUs (e.g., NVIDIA H20) and
a \emph{training pool} of high-performance, compute-optimized GPUs (e.g., NVIDIA H800). By
aligning hardware capabilities with specific phase characteristics,
disaggregation resolves the resource mismatches inherent in a monolithic
setup. Consequently, it achieves superior cost efficiency and
comparable throughput despite the introduction of cross-cluster synchronization
overheads~\cite
{zhong2025streamrl,fu2025areal,wang2025seamlessflow,yan2025arealhex}.



However, disaggregation introduces a fundamental efficiency challenge caused by \textit
{dependency bubbles}. Due to the strict synchronization required by on-policy
learning, the training cluster must remain idle during rollout, and vice
versa (\autoref{fig:core_insight}-top), leading to severe cluster
underutilization. While systems such as AReaL~\cite
{fu2025areal}, StreamRL~\cite{zhong2025streamrl}, and AsyncFlow~\cite
{han2025asyncflow} attempt to eliminate these bubbles by adopting \emph
{asynchronous, off-policy algorithms}, they achieve high utilization only by
relaxing the synchronization requirements. This relaxation introduces \emph
{sample staleness} that often compromises model accuracy and convergence
stability, rendering it unsuitable for tasks that demand strict on-policy
performance.



We present \SystemName{}, a cross-cluster scheduling framework for disaggregated RL
post-training that mitigates dependency bubbles by elevating the optimization
scope from the individual job to \emph{cluster-level orchestration}. Our key
insight is that the dependency bubbles inherent to one job can be effectively
utilized to serve another. \SystemName{} exploits this by orchestrating
multiple RL jobs into a \emph{co-execution group}, tightly ``weaving'' their
rollout and training phases across the two resource pools (\autoref
{fig:core_insight}-bottom). This \emph{co-scheduling} approach enables
efficient \emph{time-multiplexing} of both rollout and training GPUs,
maximizing utilization while preserving the synchronous dependencies required 
for on-policy learning.

While intuitive, realizing this co-scheduling benefit in production is
non-trivial due to three primary challenges. 
(\textbf{C1}) First,
production RL jobs exhibit \emph{extreme workload heterogeneity}
in model sizes (3B–32B), response lengths, and interaction patterns (\autoref
{fig:job_diversity}), leading to highly diverse phase durations and resource
demands. Naive time-multiplexing of such diverse workloads results in severe interference: for instance, pairing two rollout-heavy jobs creates a bottleneck that substantially delays both (\autoref{fig:bad_schedule}). Identifying an optimal,
interference-free schedule for these heterogeneous workloads formulates a 
Job Shop Scheduling problem~\cite{jain1999deterministic}, which is known to
be NP-hard. 
(\textbf{C2}) Second, unlike standard deep learning workloads with stable
iteration times, RL rollouts are \emph{highly stochastic}: LLM generation
follows a \emph{long-tailed distribution}~\cite{zhong2024optimizing,gao2025rollpacker,he2025history}, where a few straggler requests can
\emph{unpredictably prolong} phase durations, rendering static plans obsolete.
(\textbf{C3}) Third, 
efficient time-multiplexing is fundamentally constrained by context-switching costs.
RL post-training is inherently \emph{stateful},
requiring the management of hundreds of gigabytes of model weights and optimizer states (\autoref{tab:memory_footprint}).
Repeatedly loading these massive states over a bandwidth-limited cross-cluster
network induces prohibitive \emph{cold-start latencies}---up to 80 seconds per
switch (\autoref{fig:cold_warm_times})---which can easily offset 
the throughput gains of co-scheduling.


\SystemName{} addresses these challenges via a holistic algorithm-system
co-design. At the core is a \emph{near-optimal} scheduling algorithm. The goal
is to \emph{minimize the total resource provisioning cost}---thereby
minimizing dependency bubbles---while adhering to job-specified SLOs, defined
as the acceptable slowdown relative to \emph{solo execution} (\autoref
{fig:core_insight}-top). To tackle the intractability of heterogeneous
scheduling (\textbf{C1}), \SystemName{} decomposes the global optimization
problem into two tractable decisions: (1) \emph{inter-group scheduling}, which
identifies jobs for group co-execution and, (2) \emph{intra-group scheduling},
which orchestrates execution sequences within a co-execution group. When a
new job arrives, the \emph{inter-group scheduler} scans for an existing
co-execution group where the job can be placed without violating the SLOs of
any group member. Among all SLO-compliant placement options, it selects the one
that yields the \emph{minimum marginal provisioning cost}; if no such group
exists, it provisions a \emph{new, isolated group}. Within each group,
the \emph{intra-group scheduler} employs a round-robin schedule, a policy we
prove is \emph{optimal} for minimizing dependency bubbles in this context
(\S\ref{subsec:intra-group}).

To handle runtime stochasticity (\textbf{C2}), \SystemName{} adopts a
two-pronged strategy combining \emph{conservative admission control} with
\emph{long-tail migration}. For inter-group placement, the scheduler assumes a
worst-case scenario where all responses reach the maximum token length,
ensuring that SLO guarantees hold even under maximum load (\S\ref{subsec:inter-group}). At runtime,
the intra-group scheduler dynamically adapts to the observed response
distribution. It employs \emph{long-tail migration}, opportunistically moving
long responses to a small subset of devices to free up the majority of
the rollout pool, thereby allowing the next job to begin pipelined execution
immediately (\S\ref{subsec:intra-group})

Finally, to mitigate the prohibitive switching overheads (\textbf
{C3}), \SystemName{} implements a \emph{warm-start} mechanism. By
rightsizing co-execution groups to fit within the host memory of the worker
nodes, \SystemName{} ensures that all necessary job states, such as model
weights, optimizer states, and execution contexts, remain \emph{cached} in
host memory. When a context switch is required, the worker simply loads the
cached state from host memory to the GPU, rather than fetching it across the
slow cluster interconnect or from disk. This optimization reduces context
switching latency by two orders of magnitude (\autoref
{fig:cold_warm_times}), making fine-grained time-multiplexing practical.


To enforce these fine-grained schedules, \SystemName{} introduces a \emph
{phase-centric control model} that treats individual RL phases as first-class
schedulable entities. This abstraction exposes the job's internal dependency graph to the scheduler, transparently managing job state loading required for the
warm-start mechanism. Furthermore, \SystemName{} optimizes cross-cluster model
synchronization via a \emph{topology-aware} broadcast scheme. 
It pipes a single model copy across the slow inter-cluster link through 
parallel point-to-point streams, then utilizes high-speed local fabrics 
(e.g., NVLink or InfiniBand) for intra-cluster broadcasting, effectively eliminating the synchronization bottleneck in disaggregated setups.



We implemented \SystemName{}  atop ROLL~\cite{wang2025ROLL} and evaluated it
in a production-scale disaggregated testbed comprising a rollout pool of
328 H20 GPUs and a training pool of 328 H800 GPUs. End-to-end replays of
production workloads on these two clusters reveal that \SystemName{} reduces
total resource provisioning costs by up to 1.84$\times$ compared to naive
disaggregation and 1.38$\times$ compared to the state-of-the-art veRL~\cite
{sheng2025hybridflow} baseline, all while maintaining 100\% SLO attainment
(\S\ref{eval:end2end}). Large-scale trace-driven simulations further 
confirm that \SystemName{}'s inter- and intra-group scheduling combined operates within 6\% of the theoretical optimum identified via brute-force search 
(\S\ref{eval:global_sched}).


\section{Background and Motivation}
\label{sec:background}

\PHB{RL Post-Training Workload Characterization.}
RL post-training has evolved into a cornerstone
workload for modern AI infrastructure. In our production clusters, the volume
of RL jobs nearly tripled within six months, growing from 5k monthly jobs
in April to over 14k in September 2025, driven largely by the need to
instill complex reasoning capabilities in LLMs~\cite
{shao2024deepseekmath,deepseekr1,qwq32,wang2025ROLL,feng2025retool}. Unlike
LLM pre-training, which is a uniform compute stream, the standard RL
post-training workflow comprises repeated cycles across three phases with
distinct resource bottlenecks.
(1) \textbf{Rollout:} The actor LLM generates responses for a batch of input
prompts, which are subsequently evaluated to collect the reward feedback.
This phase is characterized by \emph{high memory bandwidth pressure} due to KV-cache
operations but relatively \emph{low arithmetic intensity}. On high-end training GPUs, 
this results in severe compute underutilization~\cite{he2025history,zhong2025streamrl,fu2025areal}. (2) \textbf
{Training:} The actor LLM's parameters are optimized based on the reward
feedback. This phase is highly \emph{compute-intensive}
and requires massive floating-point throughput and high-bandwidth interconnects
(e.g., NVLink and InfiniBand) for gradient aggregation~\cite{shoeybi2019megatron}.
(3) \textbf{Synchronization}: Updated parameters must be broadcast from training workers to rollout workers. This phase is \emph{network-bound}, often becoming a bottleneck when workers are distributed across different physical domains.

\PHM{The Case for Disaggregated RL.}
The divergent resource requirements of rollout and training create a
fundamental inefficiency in traditional \emph{monolithic, co-located deployments}~\cite
{sheng2025hybridflow}, where rollout and training are 
time-multiplexed on a single cluster of homogeneous, compute-optimized GPUs
(e.g., NVIDIA H100/H800). This forces the memory-bound rollout to run
on expensive high-FLOPS hardware, leading to significant resource
mismatches and increased total cost of ownership (TCO)~\cite
{yan2025arealhex,zhong2025streamrl}.

Disaggregation offers a promising solution to address this mismatch~\cite
{zhong2025streamrl,fu2025areal,wang2025seamlessflow,yan2025arealhex}. In this
setup, the RL workload is disaggregated across two purpose-specific resource
pools (\autoref{fig:core_insight}-bottom): training is retained on a pool 
of costly, high-FLOPS GPUs (e.g., H100/H800), while rollout is offloaded to a cluster
of cost-effective, inference-optimized GPUs (e.g., H20), which offer high HBM
capacity and bandwidth at only a fraction of the cost (\autoref{tab:gpu_price}). 
Compared to monolithic provisioning, disaggregation aligns
hardware capabilities with phase characteristics, offering
superior TCO \emph{in theory}.

\PHM{Dependency Bubbles.}
While disaggregation addresses hardware mismatches, its efficiency can be
significantly undermined by \emph{dependency bubbles}. 
State-of-the-art RL post-training relies on \emph{synchronous, on-policy} algorithms to ensure training stability and model quality~\cite{sync_rl_stable,shao2024deepseekmath,rastogi2025magistral,deepseekr1}. This synchronization constraint mandates that the training pool must remain idle while waiting for the rollout pool to generate fresh experiences, and conversely, the rollout pool must remain idle while waiting for the training pool to update parameters, as illustrated in 
\autoref{fig:core_insight}-top.

Existing systems such as AReaL~\cite
{fu2025areal,yan2025arealhex}, StreamRL~\cite{zhong2025streamrl}, and
AsyncFlow~\cite{han2025asyncflow} attempt to eliminate these bubbles by
adopting \emph{asynchronous, off-policy} algorithms. However, decoupling
rollout from training introduces \emph{sample staleness}, which frequently
compromises model convergence and final accuracy, rendering these solutions
unsuitable for tasks demanding strict on-policy performance. Consequently,
production deployments often revert to synchronous execution at the cost
of significant resource idleness. In fact, our evaluation in \S\ref{eval:end2end}
reveals that the idle time induced by dependency bubbles forces disaggregated 
setups to incur \emph{even higher provisioning costs} (\$0.94k/h) than the 
hardware-mismatched co-located baselines (\$0.71k/h), despite using cheaper GPUs 
for rollout. Thus, without a scheduling mechanism to reclaim this lost capacity, 
the theoretical TCO benefits of disaggregation are effectively nullified by 
system-level inefficiencies.

\begin{table}[tb]
    \centering
    \footnotesize
    \begin{tabular}{ccccc}
        \toprule
         \multirow{2}{*}{\textbf{Accelerator}} & \textbf{Comp.} & \textbf{HBM Cap.} & \textbf{HBM B/w} & \textbf{Cost} \\
         & \textbf{(TFLOPS)} & \textbf{(GB)} & \textbf{(TB/s)} & \textbf{(\$/h)~\cite{zhu2025megascaleinfer}} \\
         \midrule
         H20 & 148 & 96 & 4.0 & 1.85\\
         H800 & 989.5 & 80 & 3.35 & 5.28\\
         \bottomrule
    \end{tabular}
    \caption{Performance specifications and cost-effectiveness
of the GPUs used in our disaggregated clusters~\cite{zhu2025megascaleinfer, zhong2025streamrl}. }
    \label{tab:gpu_price}
\end{table}
\section{Scheduling Opportunities and Challenges}
\label{sec:challenges}

In this section, we identify the opportunities for mitigating dependency
bubbles in disaggregated RL post-training through cluster-level scheduling.
We then examine the algorithmic and system-level challenges in
realizing these opportunities.


\subsection{The Co-Scheduling Opportunity}
\label{subsec:potential}

From the perspective of a single job, the dependency bubbles described
in \S\ref{sec:background} are unavoidable without violating the
synchronization requirements of on-policy algorithms. However, in
\emph{shared, multi-tenant} clusters running diverse RL workloads, these individual
 inefficiencies represent available capacity that can be reclaimed through
 global orchestration.

Our key insight is that the idle resources constituting one job's dependency
bubbles can be utilized to execute the active phase of another. By orchestrating
jobs into \emph{co-execution groups}, the scheduler can tightly ``weave''
together their workflows, ensuring that the compute-intensive training phase
of one job executes in parallel with the memory-bound rollout phase of
another (\autoref{fig:core_insight}-bottom). This interleaved execution
pattern effectively hides dependency bubbles, allowing the system to
simultaneously saturate both the cost-effective rollout pool and the
high-performance training pool, thereby maximizing cluster-wide efficiency
without compromising the synchronization requirements of on-policy learning.



\subsection{Challenges}
\label{subsec:challenges}

However, fully unlocking the benefits of co-scheduling at production scale
is non-trivial due to three primary challenges.


\begin{figure}[tb]
    \centering
    \includegraphics[width=0.9\linewidth]{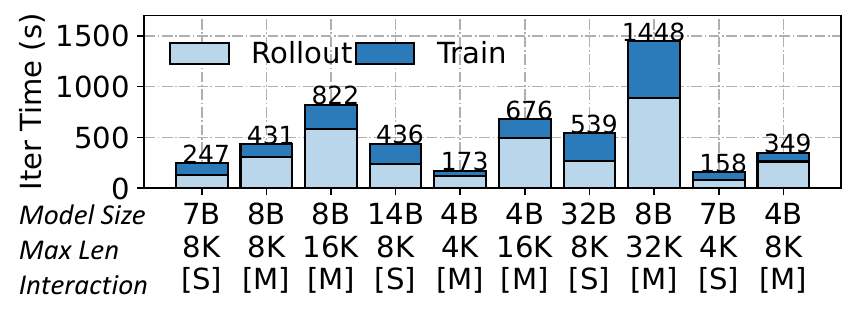}
    \caption{Top 10 popular RL post-training workloads in our production cluster: jobs' phase durations are highly diverse. [S], [M] refers to single/multi-turn interaction during rollout.}
    \label{fig:job_diversity}
\end{figure}

\begin{figure}[tb]
    \centering
    \includegraphics[width=0.9\linewidth, trim=0 10 0 5, clip]{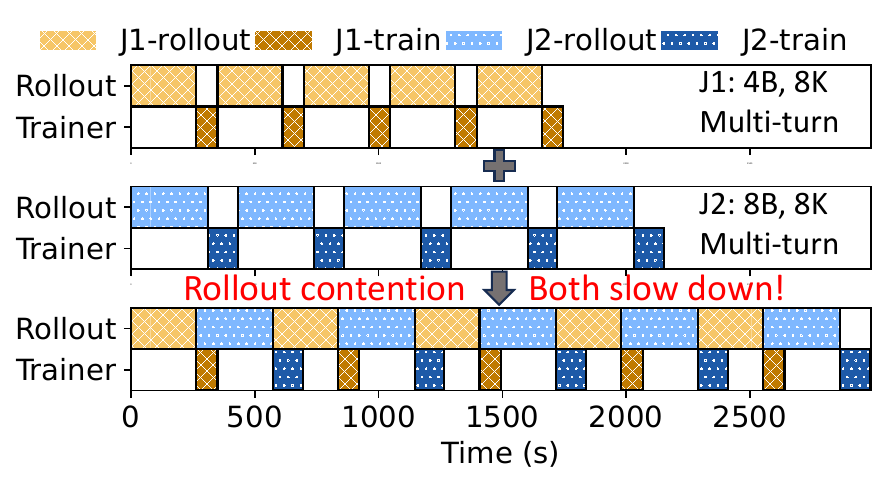}
    \caption{A bad case of naive time-multiplexing: two rollout-heavy jobs compete for a rollout node and both slow down.}
    \label{fig:bad_schedule}
\end{figure}

\PHM{C1: Workload Heterogeneity and Scheduling Intractability.}
First, realizing efficient co-scheduling is complicated by the extreme
diversity of production RL workloads. As depicted in \autoref
{fig:job_diversity}, jobs in our cluster span a wide spectrum of model sizes
(3B--32B), response lengths (4k--32k tokens), and interaction modes
(single- vs. multi-turn). This heterogeneity manifests as highly variable
phase durations, ranging from 50s to over 900s, and significant \emph
{phase skew}; for instance, multi-turn agentic workloads often exhibit
rollout phases that are 3$\times$ to 4$\times$ longer than their corresponding training
phases.

Consequently, naive time-multiplexing often proves detrimental due to resource
contention. For example, arbitrarily pairing two rollout-heavy jobs creates a
bottleneck on the inference nodes, forcing both jobs to stall. As illustrated
in \autoref{fig:bad_schedule}, such contention results in severe performance degradation,
slowing down concurrent jobs by 1.40$\times$ and 1.64$\times$, respectively. To avoid such
interference, the scheduler must identify optimal packings that satisfy
strict performance SLOs. However, mapping these heterogeneous, phase-skewed
workloads to available resources formulates a Job Shop Scheduling problem~\cite
{jain1999deterministic,garey2002computers},
which is known to be NP-hard even under the simplifying assumption of
deterministic phase durations.



\PHM{C2: Stochastic Runtime and Skewness Bubbles.}
Second, efficient scheduling orchestration is complicated by the
inherent \emph{stochasticity} of RL workloads. Unlike pre-training, RL
rollout creates a dynamic workload where execution time depends on the
variable length of generated responses, which follows a \emph
{long-tail distribution} (\autoref
{fig:eval_migration}). This introduces two distinct system challenges. 
First, the workload is \emph{non-stationary}: the distribution of response lengths 
drifts across iterations as the model updates, with a few "straggler" requests frequently reaching the maximum token limit~\cite
{zhong2024optimizing,gao2025rollpacker,zhong2025streamrl}. 
Because training computation scales linearly with token count,
this rollout variance propagates to the subsequent training phase, rendering static
orchestration planning obsolete. Second, long-tail responses 
result in \emph{skewness bubbles} during the rollout phase~\cite{zhong2025streamrl,zhong2024optimizing}. Within a rollout batch, early-finishing GPUs must idle while
waiting for stragglers, effectively serializing the batch completion. This
forces the scheduler to solve a dynamic variant of the Job Shop Scheduling
problem, where task durations are unpredictable, time-varying, and prone to
significant internal resource fragmentation.




\begin{table}[t]\footnotesize
  \centering
  \begin{tabular}{lcccc}
    \toprule
    \textbf{Model Size} & \textbf{3B} & \textbf{7B} & \textbf{14B} & \textbf{32B} \\
    \midrule
    Rollout & 113.4 & 275.7 & 445.4 & 490.3(TP=2) \\
    Train & 156.2 & 240.0(TP=2) & 456.1(TP=2) & 520.4(TP=4) \\
    \bottomrule
  \end{tabular}
  \caption{Memory footprint (GB) required for caching rollout or training actors on an 8-GPU node across model sizes.}
  \label{tab:memory_footprint}
\end{table}

\begin{figure}[tb]
    \centering
    \includegraphics[width=0.95\linewidth]{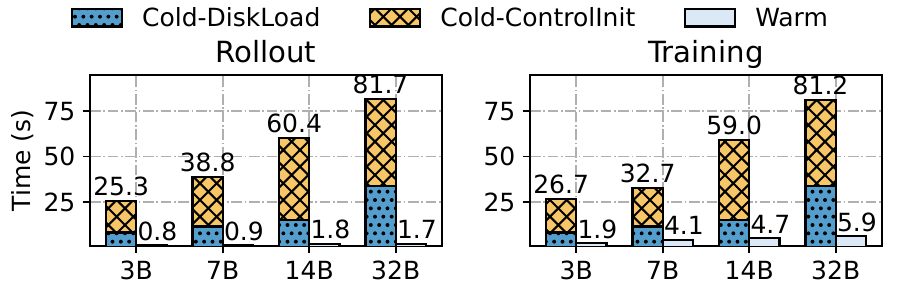}
    \caption{Cold and warm start latency for rollout (\textbf{left}) and training (\textbf{right}) across model sizes on an 8-GPU node.}
    \label{fig:cold_warm_times}
\end{figure}

\PHM{C3: Context Switching Overhead and Memory Residency.}
Third, the granularity of efficient time-multiplexing is fundamentally constrained by the cost of inter-job context switching. Unlike stateless LLM inference~\cite{fu2024serverlessllm}, RL post-training is inherently \emph{stateful}, managing a massive working set that includes large model weights, optimizer states, and complex execution contexts such as dataset pipelines and environment states.
Reconstructing these states from scratch upon every switch---known as a \emph{cold start}---incurs prohibitive overheads due to both data- and control-plane re-initialization. As shown in \autoref{fig:cold_warm_times}, cold-starting a rollout or training phase on an 8-GPU node (H20 for rollout and H800 for training) takes up to 80 seconds, degrading end-to-end throughput by up to 45\%.


Furthermore, unlike
serverless systems that can mitigate cold starts via high-speed RDMA state
transfers~\cite
{zhang2025blitzscale,yu2025lambda,fu2024serverlessllm}, disaggregated RL
setups are bottlenecked by limited cross-cluster Ethernet bandwidth.
Consequently, the only viable mechanism for rapid switching is a \emph
{warm-start} strategy, where job states remain cached in local host DRAM.
While this approach significantly reduces switching latency by up to
48$\times$ (\autoref{fig:cold_warm_times}), it imposes severe memory
pressure. Since a single phase's state consumes hundreds of gigabytes
(\autoref{tab:memory_footprint}), even high-memory nodes (1--2 TB) are
strictly limited to a residency of two to five concurrent jobs. This creates a tight \emph{residency constraint}, forcing the scheduler to optimize for utilization within a bounded memory budget.

\section{The \SystemName{} Scheduling Design}
\label{sec:algorithm_design}

We present \SystemName{}, a cluster scheduling framework that reclaims
dependency bubbles for disaggregated RL post-training through cross-cluster
orchestration. \SystemName{} adopts a holistic algorithm-system co-design: we
decouple the logical scheduling policy (\S\ref{sec:algorithm_design}) from
the underlying execution plane (\S\ref{sec:exec_plane}). This section details
the core scheduling algorithms.

\subsection{Co-Execution Group}
\label{subsec:coexecgroups}

\SystemName{}'s scheduling objective is to minimize total resource
provisioning cost---and thereby minimize dependency bubbles---while strictly
adhering to job performance SLOs and node memory constraints. To achieve
this, we introduce the \emph{co-execution group} abstraction. A co-execution
group is a set of jobs that \emph{share} a specific pair of rollout and
training resource pools via \emph{time-multiplexing}. Within a group, all
rollout phases execute on the group's assigned rollout workers, and all
training phases execute on the group's training workers. By partitioning jobs
into \emph{disjoint groups}, \SystemName{} transforms the intractable
global co-scheduling problem into a collection of independent,
parallel sub-problems within groups. 

This group-based scheduling directly addresses two primary challenges
identified in~\S\ref{subsec:challenges}. First, by decomposing the
cluster-wide search space into smaller, isolated groups, \SystemName
{} ensures that scheduling decisions remain \emph{computationally tractable} at
production scale, even with thousands of concurrent jobs (\textbf{C1}). 
Second, the group abstraction serves as a
strict \emph{locality domain}. By pinning jobs to specific sets of
nodes, \SystemName{}  enforces the residency constraint: it ensures that the
massive working sets (weights and optimizer states) of all group members
remain resident in host DRAM. This guarantees that context switches can be
served via high-speed local memory transfers (warm starts) rather than slow
disk or cross-cluster fetches (\textbf{C3}).

The co-execution group abstraction naturally leads to a \emph
{two-tier scheduling hierarchy}: (1) \textit{inter-group scheduling} (\S\ref
{subsec:inter-group}), which assigns arriving jobs to groups to minimize 
provisioning costs while satisfying memory and SLO constraints, and (2) \textit
{intra-group scheduling} (\S\ref{subsec:intra-group}), which orchestrates
the runtime execution order of job phases within a group to minimize
dependency bubbles.

\vspace{-.1in}

\subsection{Inter-Group Scheduler}
\label{subsec:inter-group}


\PHB{Problem Formulation.} 
We model the cluster as a collection of disjoint co-execution groups. We
define a co-execution group $G$ as a tuple $(J_G, R_{G}, T_{G}, \Phi_G)$, 
where $J_G$ is the set of active
RL jobs in the group, $R_G$ and $T_G$ denote the sets of
rollout (e.g., H20) and training (e.g., H800) GPUs provisioned for the group,
and $\Phi_G = \{P_{j}\}_{j\in J_G}$ is the set of resource placements,
where $P_j$ specifies the exact subset of rollout and training nodes job $j$ is pinned to.
This pinning $P_{j}$ strictly determines where the job's
state is cached to enable its warm start.

The inter-group scheduler solves the following \emph{online placement
problem}: upon the arrival of a job $j$, it must assign $j$ to a co-execution
group---either an existing one or a newly created one---and allocates
specific resource placement $P_{j}$. To make optimal placement, we define
the provisioning cost of a group, $\text{Cost}(G)$, as the 
aggregate hourly price (\autoref{tab:gpu_price}) of all allocated GPUs in its rollout and
training pools ($R_{G}$ and $T_{G}$). The scheduler's objective is to
minimize the \emph{marginal provisioning cost} $\Delta$ incurred by admitting job
$j$:
\[
    \textstyle
    \min_G \Delta = \text{Cost}(G') - \text{Cost}(G),
\]
where $G'$ represents the group's state after accommodating job $j$. This
formulation naturally encourages ``packing'' jobs into existing dependency bubbles 
(where $\Delta = 0$) over provisioning new hardware (where $\Delta > 0$).
The placement decision is subject to two critical constraints:

\textbf{\emph{1) Memory Residency.}}
To guarantee warm starts (\textbf{C3}), the aggregate working set of all jobs 
pinned to a specific node must not exceed that node's host memory capacity.

\textbf{\emph{2) SLO Attainment.}} The placement must satisfy the 
performance SLOs of both the new job and all existing jobs. The SLO is defined
by each job as the \emph{tolerance for co-execution slowdown} (e.g., $1.1\times$) 
relative to solo execution.\footnote{We assume a tight SLO, e.g., tolerance for up to $2\times$ slowdown.} 
Formally, for every job $k$ in the updated group $G$, we require:
\[
    T^{\text{co-exec}}_k \le \text{SLO}_{k} \times T^{\text{solo}}_{k}.
\]
Here, $T^{\text{solo}}_{k}$ is the estimated iteration time when job $k$ is
running alone (\autoref{fig:core_insight}-top), which is simply the sum of
its rollout and training phase durations;
$T_k^{\text{co-exec}}$ is the expected iteration time under co-execution, 
which is derived by simulating the intra-group schedule (\S\ref{subsec:intra-group}). 



\PHM{Making Placement Decisions.}
Navigating the search space to find an optimal placement is non-trivial
due to both workload heterogeneity (\textbf{C1}) and runtime stochasticity
(\textbf{C2}). \SystemName addresses these complexities with three strategies.

\textbf{\emph{1) Handling Stochasticity via Conservative Planning.}} 
To guarantee SLO compliance despite the volatile, unpredictable job execution
time (\textbf{C2}), \SystemName{} decouples admission control from runtime
optimization. The inter-group scheduler acts as a ``gatekeeper'' that makes
placement decisions based on \emph{worst-case execution bounds}.
Specifically, for an arriving job $j$, we estimate its phase durations ($T_{j}^{\text{roll}}$ and $T_{j}^{\text{train}}$) assuming that every
generated response reaches the \emph{maximum token limit} defined in the job
configuration. By planning against this upper bound, we ensure that the
chosen placement satisfies the SLO constraints even under the most adverse
stochastic conditions. If the actual runtime durations are shorter, which is typical,
the intra-group scheduler dynamically reclaims the resulting slack to improve 
utilization (\S\ref{subsec:intra-group}).

\begin{figure}[tb]
    \centering
    \includegraphics[width=0.9\linewidth]{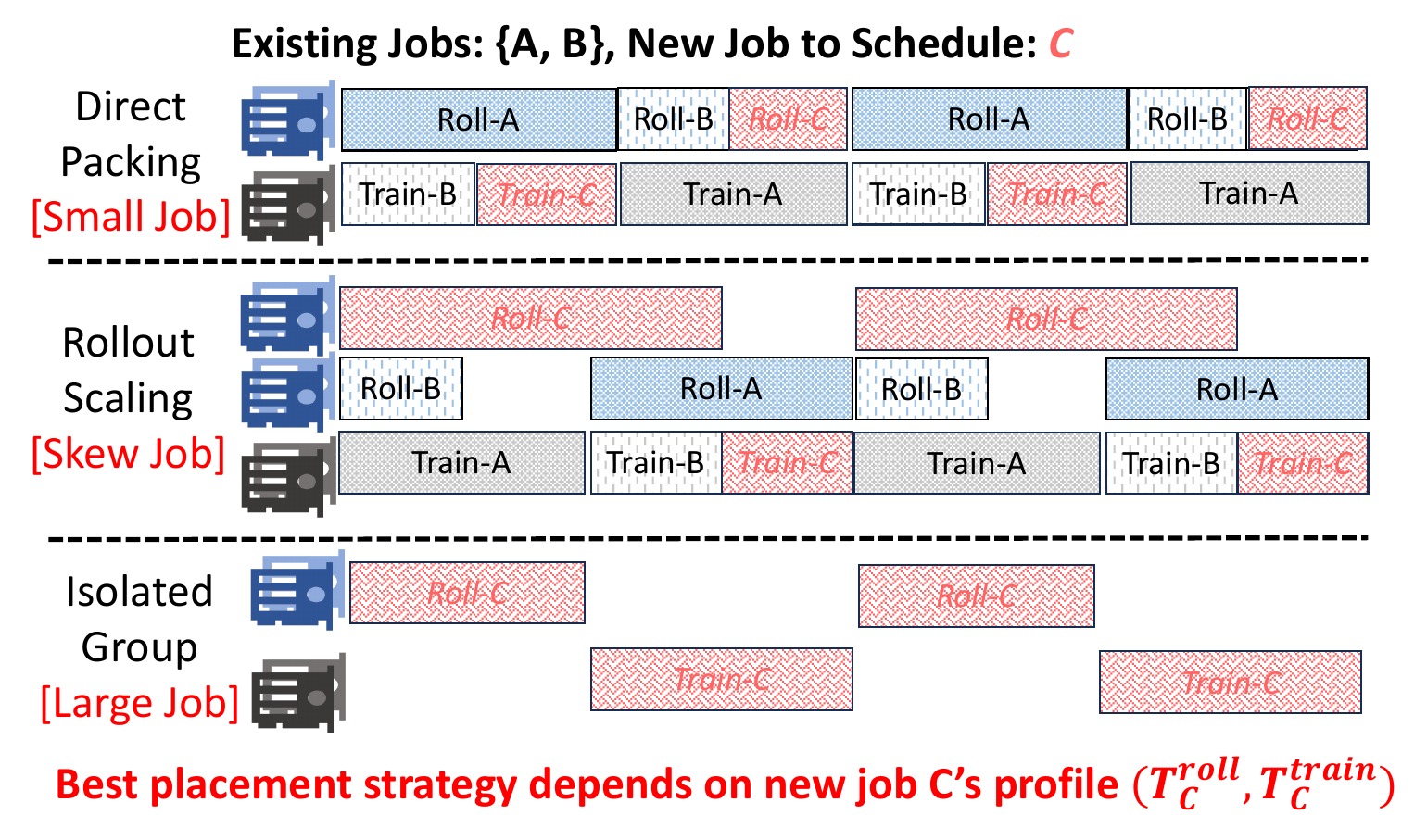}
    \caption{Placement strategies of the inter-group scheduler.}
    \label{fig:placement_strategy}
\end{figure}

\textbf{\emph{2) Optimal Placement Search.}} 
With these conservative estimates, \SystemName{} performs a global search to
minimize the marginal provisioning cost $\Delta$. For each arriving job, the
scheduler iterates through all candidate groups and evaluates three
placement strategies:
\begin{itemize}[topsep=3pt, leftmargin=1.5em, noitemsep, nolistsep, parsep=2pt, partopsep=0pt]
    \item \underline{\emph{Direct Packing:}} 
    Inserting the job into existing dependency bubbles within a group without 
    provisioning new resources (\autoref{fig:placement_strategy}-top). This maximizes 
    utilization of already-paid-for capacity.
    \item \underline{\emph{Rollout Scaling:}} If a group has available training capacity but 
    is bottlenecked on inference, which is common with rollout-heavy agentic workloads, \SystemName{}
    scales up the group's rollout pool by provisioning \emph{just enough} rollout nodes
    to accommodate the new job (\autoref{fig:placement_strategy}-middle). 
    \item \underline{\emph{Isolated Provisioning:}} As a fallback, \SystemName{} 
    provisions a new, isolated group for the new job (\autoref{fig:placement_strategy}-bottom). 
\end{itemize}
The scheduler iterates through these strategies\footnote{To avoid the
significant overhead of reconfiguring distributed parallel groups, 
\SystemName does not scale the training pool but simply adjusts the arriving job's data parallelism 
degree to match the training pool size.}, selecting the \emph{valid placement}
that minimizes the marginal cost $\Delta$ while satisfying both the memory residency
and SLO constraints.

\begin{figure}[tb]
    \centering
    \includegraphics[width=\linewidth]{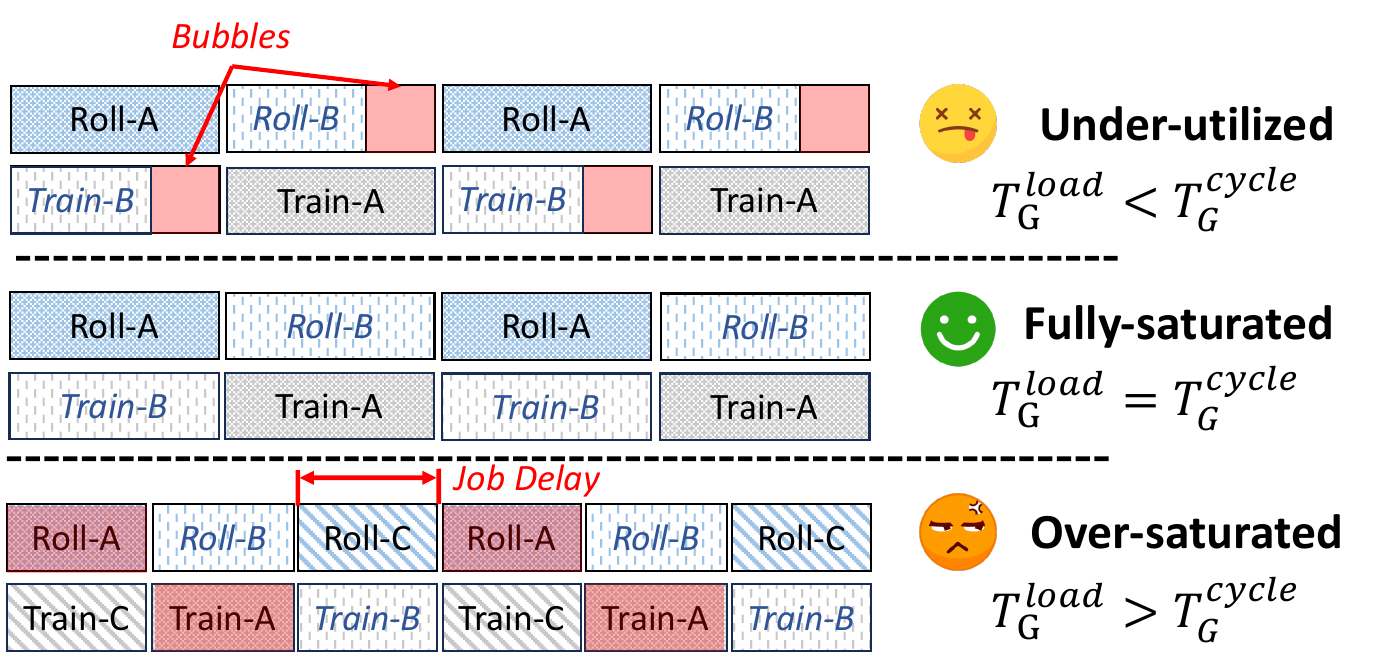}
    \caption{Status of a co-execution group. \SystemName only places new jobs into under-utilized groups with unsaturated dependency bubbles and avoids creating over-saturated groups.}
    \label{fig:group_status}
\end{figure}

\textbf{\emph{3) Pruning Saturated Groups.}} 
To ensure this search remains tractable at production scale (\textbf{C1}), 
\SystemName{} proactively prunes the search space. Before evaluating specific 
placements, the scheduler filters out groups that are already \emph{saturated},
where the aggregate job load reaches the group's bottleneck resource
capacity, and adding more work to the group would lead to performance degradation.

Formally, for a group $G$, 
let $T_G^\text{cycle} = \max_{j \in J_G} T^{\text{solo}}_{j}$ be
the natural cycle iteration time dictated by the \emph{longest job} in the group.
We define the group's bottleneck load $T_G^{\text{load}}$ as the total time
required to process all phases on the bottleneck node.
Since all training nodes have identical phases, while rollout nodes may differ (see \autoref{fig:placement_strategy}), we have
\[
  T_G^{\text{load}} = \max\left(\sum_{j \in J_G} T^{\text{train}}_{j}, \max_{n\in \Phi_G}\left(\sum_{j \text{ on node } n} T^{\text{roll}}_{j}\right)\right).
\]



If $T_G^{\text{cycle}} \ge T_G^{\text{load}}$, the group is saturated,
containing no ``slack'' to absorb new work (\autoref
{fig:group_status}). Such groups are pruned immediately as any
further addition would force delays.


\PHM{Algorithm Summary.} 
We integrate these strategies into a unified online scheduling logic detailed
in \autoref{algo:inter_scheduling}. The algorithm takes a new job $j$ and the
set of existing groups as input. To find the optimal placement, the algorithm
first iterates through all existing groups (line~\ref{line:enum}), discarding
those that are already saturated (line~\ref{line:skip}). For each remaining
candidate group, it evaluates potential placement strategies for the job
(direct packing or rollout scaling); placements that would violate memory
constraints (line~\ref{line:mem}) or SLO constraints (line~\ref
{line:slo}) are discarded. The algorithm evaluates the marginal cost for each
feasible placement (lines \ref{line:placement_gen}--\ref
{line:calc_delta}) and updates its records if the placement leads to a lower
cost (lines \ref{line:upd_begin}--\ref{line:upd_end}). Finally, the algorithm
compares its records against the baseline cost of provisioning a
fresh, isolated group (lines \ref{line:solo_begin}–\ref{line:solo_end}) and
returns the group and job placement that yield the lowest costs.

The algorithm allows for highly efficient decision-making. Since the number of 
placement strategies per group is small, the search complexity is 
\emph{linear} with respect to the number of active groups.
As empirically demonstrated in \S\ref{eval:global_sched}, this heuristic 
allows the scheduler to make optimal decisions in sub-seconds 
even in clusters with thousands of jobs.

\begin{algorithm}
\footnotesize
\caption{Inter-Group Scheduling Algorithm}
\label{algo:inter_scheduling}
\begin{algorithmic}[1]
\Require Job to schedule $j$, all existing groups $\{G_i\}_{i=1}^n$.
\Ensure Best group $G^*$, best placement $P_j^*$.
\Procedure{Schedule}{$j, \{G_i\}_{i=1}^n$}
    \State $\Delta^* \gets \infty$, $G^* \gets \text{None}$, $P_j^* \gets \text{None}$
    \Comment{\textcolor{blue}{Initialize best values.}}
    \For{each group $G$ in $\{G_i\}_{i=1}^n$}\label{line:enum} \Comment{\textcolor{blue}{Try all existing groups.}}
        \If{$T_G^{\text{load}}\ge T_G^{\text{cycle}}$} \Comment{\textcolor{blue}{Skip saturated groups.}} \label{line:skip}
            \State \textbf{continue}
        \EndIf
        \State $\mathcal{P}\gets \textsc{GeneratePlacements(G)}$\label{line:placement_gen}
        \For{each resource placement $P_j$ in $\mathcal{P}$}
            \If{$j.\text{mem\_req} \ge \min_{\text{node}\in P_j}(\text{node}.\text{mem\_avail})$}\label{line:mem}
                \State \textbf{continue}
            \EndIf
            \If{exists $k\in \{j\}\cup J_G$, s.t., $T^{\text{co-exec}}_{k} > \text{SLO}_k \times T^{\text{solo}}_{k}$}\label{line:slo}
                \State \textbf{continue}
            \EndIf
            \State $\Delta \gets \text{Cost}(G\cup\{(j, P_j)\}) - \text{Cost}(G)$ \label{line:calc_delta}
            \If{$\Delta< \Delta^*$}\label{line:upd_begin}
                \State \textbf{Update} $\Delta^* \gets \Delta$, $G^* \gets G$, $P_j^* \gets P_j$
            \EndIf\label{line:upd_end}
        \EndFor
    \EndFor
    \State $\Delta \gets \text{Cost}(\{j\}, \{\})$ \label{line:solo_begin}
    \Comment{\textcolor{blue}{Try to place $j$ in a new group.}}
    \If{$\Delta < \Delta^*$}
        \State \textbf{Update} $\Delta^* \gets \Delta$, $G^* \gets \{j\}$, $P_j^* \gets \{\}$
    \EndIf \label{line:solo_end}
    \State \Return $G^*, P_j^*$
\EndProcedure
\end{algorithmic}
\end{algorithm}

\vspace{-.2in}
\subsection{Intra-Group Scheduler}
\label{subsec:intra-group}

Once the inter-group scheduler assigns a job to a group, the intra-group
scheduler is responsible for orchestrating the runtime execution sequence.
Its primary objective is to maximize resource utilization---and thereby
minimize dependency bubbles---within the assigned resource pools.

\PHM{The Round-Robin Policy.}
\SystemName{} employs a \emph{cyclic round-robin schedule}. Within a
 co-execution group, the scheduler defines a \emph{meta-iteration} in which
 every active job executes exactly one rollout phase and one training phase
 (\autoref{fig:core_insight}).
 These phases are orchestrated sequentially on their assigned resource pools.
 For example, in a group with jobs $\{A, B\}$, the rollout pool executes
 $\text{Roll}_A \rightarrow \text{Roll}_B$, while the training pool executes
 $\text{Train}_A \rightarrow \text{Train}_B$.


While simple, this policy is \emph{optimal} under the preconditions
enforced by \SystemName. Recall that the inter-group scheduler (\S\ref
{subsec:inter-group}) proactively prunes any group where the aggregate load
exceeds the natural cycle time ($T_G^{\text{load}} > T_G^{\text
{cycle}}$). For the remaining unsaturated groups, their optimality is provable.

\begin{theorem}[Utilization Optimality]
    For any unsaturated group $G$, a meta-iteration schedule that executes 
    each job's phases exactly once in a round-robin order maximizes the aggregate 
    utilization of both rollout and training pools.
\end{theorem}

\underline{\emph{Proof Sketch.}} 
The optimality rests on the definition of an \textit{unsaturated group}: the
bottleneck node's total workload $T_G^{\text{load}}$ is no more than the
longest job's standalone cycle time $T_G^{\text{cycle}}$. Intuitively, this
implies that we can pack all other jobs' corresponding phases (e.g., their
rollout phases) into the longest job's dependency bubbles (e.g., its idle
rollout nodes during training). This ensures a round-robin cycle that
executes each job's phases exactly once to complete in time $T_G^{\text{cycle}}$. 
We then empirically show any deviation from this simple schedule is
\emph{suboptimal}. (1) \textit{Executing less is impossible}: Omitting any job from
the cycle leads to more bubbles and starvation, which is trivially
non-optimal. (2) \textit{Executing more is inefficient}: Repeating any job
phase prolongs the cycle time as the added phase can only start after
finishing the slowest job. However, this added duration is disproportionately
larger than the useful work added, leading to a net decrease in resource
utilization. 

Therefore, the round-robin schedule, which executes all required work in the shortest possible cycle time, is utilization-optimal. We provide a formal proof in Appendix~\ref{sec:appendix_proof}.





\begin{figure}[tb]
    \centering
    \includegraphics[width=0.9\linewidth]{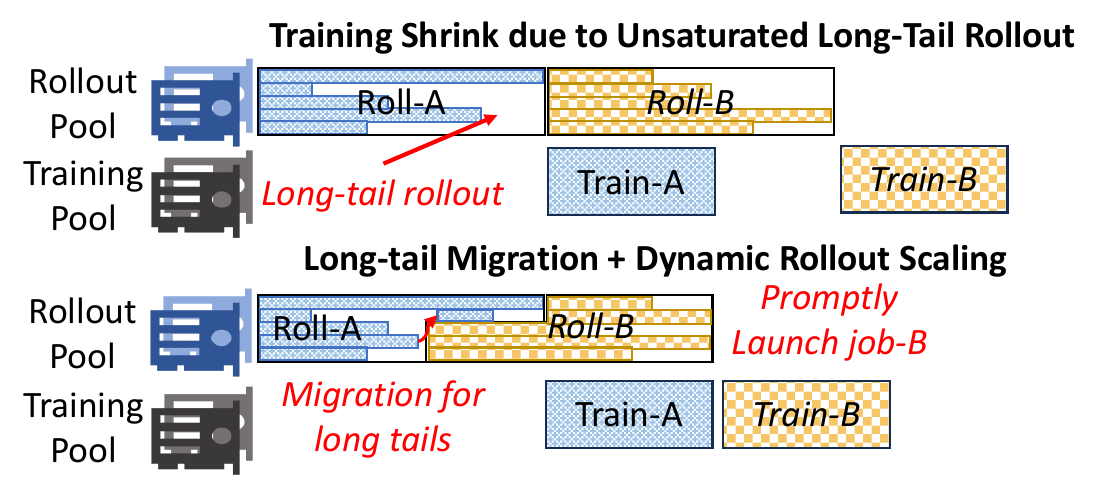}
    \caption{Long-tail migration effectively handles dynamism.}
    \label{fig:migration_design}
\end{figure}

\PHM{Long-Tail Migration.} 
While the round-robin schedule is optimal for deterministic workloads,
production RL phases are highly stochastic (\textbf{C2}). Specifically,
rollout durations follow a \emph{heavy-tailed distribution} where the
completion time of an input batch is dictated by a small number of ``straggler''
responses that reach maximum token limits~\cite
{zhong2024optimizing,zhong2025streamrl,gao2025rollpacker,he2025history}. This
phenomenon creates significant \emph{intra-phase fragmentation}: as the majority of
responses finish early, most GPUs in the rollout pool idle
wait for the few stragglers to complete, creating ``skewness bubbles'' 
(\autoref{fig:migration_design}-top).

To reclaim this fragmented capacity, \SystemName{} employs \emph
{long-tail migration} to dynamically adapt the schedule at runtime. 
The intra-group scheduler continuously
monitors the progress of active rollout phases. When a phase enters a
\emph{tail-bound state}, triggered when a threshold of responses (e.g., 80\%) have
completed, the system interrupts the execution, consolidates the remaining
long-tail responses onto a small subset of workers, and immediately
starts the \emph{next} job's rollout phase on the newly freed rollout GPUs
(\autoref{fig:migration_design}-bottom).
This mechanism effectively pipelines the tail of one job with the head of the
next, ensuring high utilization and faster completion.


\subsection{Generality and Composability}
\label{subsec:generality}

\SystemName{} is agnostic to the specific RL algorithm employed. Its
 scheduling mechanism generalizes to diverse on-policy RL algorithms,
 including PPO~\cite{schulman2017proximal}, GRPO~\cite
 {shao2024deepseekmath}, and DAPO~\cite{yu2025dapo}. While primarily designed for on-policy algorithms, \SystemName remains applicable to off-policy jobs (e.g., one-step off-policy~\cite{zhong2025streamrl}) that exhibit structural dependency bubbles due to insufficient overlap between training and rollout. Moreover, \SystemName{}'s 
 cluster-level orchestration is orthogonal to intra-job optimizations.
 Techniques such as parameter relocation~\cite{mei2024real}, request-level
 tail batching~\cite{gao2025rollpacker}, and speculative decoding~\cite
 {qin2025seer, chen2025respec, he2025history, hu2025taming} operate
 within the scope of a single job or phase, making them fully
 composable with \SystemName{}.
 \vspace{-.1in}

\section{The \SystemName Execution Plane}
\label{sec:exec_plane}


The scheduling policies described in \S\ref{sec:algorithm_design} provide a
theoretical blueprint for reclaiming dependency bubbles. However, realizing
these gains in production clusters requires an execution plane capable of 
enforcing fine-grained decisions. This section details the system mechanisms
that bridge this gap. We focus on two implementation challenges: (1) enabling
rapid context switching via a phase-centric control model
(\S\ref{subsec:phase_centric_control}), and (2) mitigating cross-cluster
bandwidth bottlenecks via topology-aware model synchronization (\S\ref
{subsec:commopt}).

\vspace{-.1in}
\subsection{Phase-Centric Control}
\label{subsec:phase_centric_control}

Conventional deep learning schedulers view the ``job'' as the atomic unit of
resource allocation. This coarse granularity is insufficient to interleave
distinct phases of different jobs on the same hardware. To address
this, \SystemName{} introduces a \emph{phase-centric execution model} that elevates
individual RL phases to first-class schedulable entities.


\PHM{Declarative Phase Management.}
We model each RL job as a dependency graph of phases. After a one-time
initialization (\texttt{Init}) of the job states (e.g., models, datasets),
the job enters a cyclic dependency loop: \texttt
{Rollout} $\rightarrow$ \texttt{Train} $\rightarrow$ \texttt
{Sync}. \SystemName exposes this internal structure to the scheduler via a
declarative Python API. Users simply annotate their phase functions with
a \texttt{@rollmux.phase} decorator, which injects a \emph{transparent runtime
shim} to manage the execution lifecycle. When a phase is invoked, this shim
first blocks execution until it acquires a run permit from the intra-group
scheduler. Upon approval, it performs a \emph{warm start} by loading the
phase's resident working set from host DRAM into GPU memory. Once the user function completes, the shim
immediately offloads the updated state back to host memory and releases the
GPU resources, making the hardware instantly available for the next phase in
the group's queue.
Crucially, \SystemName{} optimizes this switching process by decoupling data
plane state from control plane context. Naively terminating a process after a
phase completes would force the system to tear down and rebuild expensive
control plane (e.g., NCCL communicators, environment handles) upon every
switch. Instead, \SystemName{} employs a \emph{lightweight suspension}
strategy: after offloading, the shim places the process into a
sleep loop while retaining its control plane context without consuming GPU
resources. On the next wake-up, resuming the phase only requires reloading its
cached state onto the GPU, avoiding expensive cold starts
from disk and control-plane re-initialization.


\PHM{Runtime Hooks.}
Finally, the system exposes a runtime hook interface \texttt
{@rollmux.runtime\_hook}. This interface serves two critical roles.
First, it drives the round-robin schedule: the intra-group
scheduler maintains a FIFO queue for each worker node. When a job's phase
completes, the hook signals the scheduler to enqueue the job's next phase
onto the alternate resource pool's queue (e.g., moving from rollout to
training) and starts the next waiting phase on the now-idle
resources. Second, it enables the long-tail migration (\S\ref{subsec:intra-group}). By
exposing internal token generation progress, the hook
allows the scheduler to detect tail-bound states and externally trigger
migration, dynamically reconfiguring resources in real-time.
\vspace{-.1in}

\subsection{Topology-Aware Model Synchronization}
\label{subsec:commopt}

\begin{figure}[tb]
    \centering
    \includegraphics[width=0.95\linewidth]{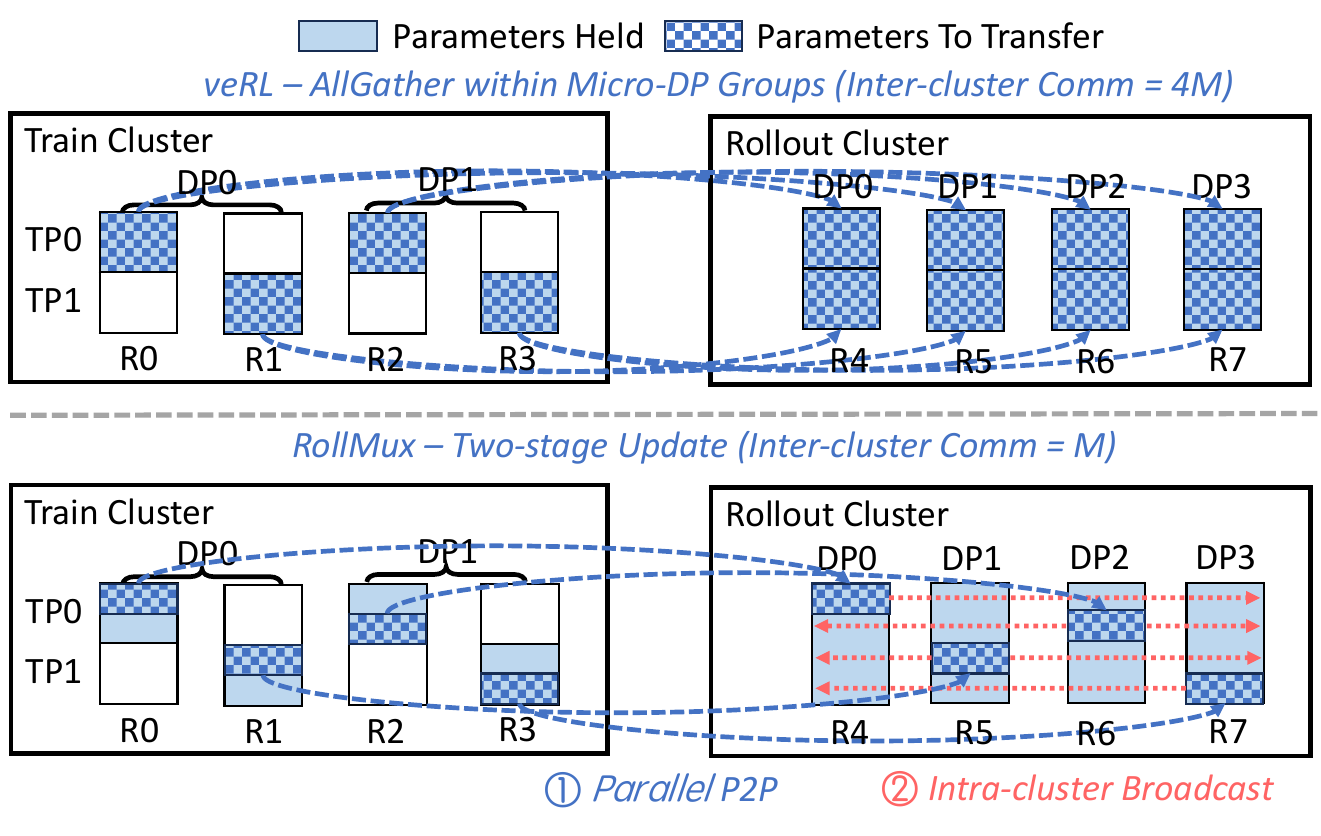}
    \caption{A synchronization example from the training cluster (TP=2, DP=2) to the rollout cluster (DP=4), where \SystemName sends exactly one copy across the cross-cluster network. $M$ denotes the number of total parameters.}
    \label{fig:disagg_update}
\end{figure}

To mitigate the cross-cluster bandwidth bottleneck, \SystemName employs a
\emph{topology-aware} communication strategy to efficiently synchronize model
parameters from the training cluster to the rollout cluster.
State-of-the-art RL frameworks (e.g., veRL~\cite{sheng2025hybridflow}) rely on flat collective operations like \texttt{AllGather} to propagate model updates. While efficient in monolithic clusters, this approach is pathological in disaggregated setups. It treats the slow cross-cluster Ethernet link and the fast intra-cluster InfiniBand/NVLink fabric as a single uniform network. Consequently, it forces every rollout worker to independently fetch a full copy of the model parameters over the slow cross-cluster link  (\autoref{fig:disagg_update}-top), causing a severe bottleneck while leaving local high-speed fabrics idle.

\SystemName{} eliminates this inefficiency by replacing the flat collective
 with a \emph{hierarchical two-stage transfer}. \textcircled{1} In the first stage
 (\textbf{inter-cluster scatter}), \SystemName{} partitions the updated model into $N$
 disjoint shards, where $N$ is the number of training GPUs. Each training
 GPU transmits a unique shard to a corresponding rollout GPU via parallel
 point-to-point (P2P) streams. This ensures that \emph{exactly one full copy} of the
 model traverses the slow cross-cluster link. \textcircled{2} In the second
 stage (\textbf{intra-cluster broadcast}), the receiving GPUs immediately disseminate
 their shards to all other rollout workers using the high-bandwidth InfiniBand/NVLink
 fabric. This two-stage pipeline effectively mitigates
 the slow cross-cluster bottleneck, minimizing overall synchronization time
 and fully utilizing network hierarchies.


\section{System Implementation}
\label{sec:impl}

We implemented \SystemName{} as a fully functional cluster scheduling
framework atop ROLL~\cite{wang2025ROLL}. The system comprises approximately
5.2k lines of code (LoC), written primarily in Python for controllers and C++ for communication modules.

\begin{figure}[tb]
    \centering
    \includegraphics[width=0.8\linewidth]{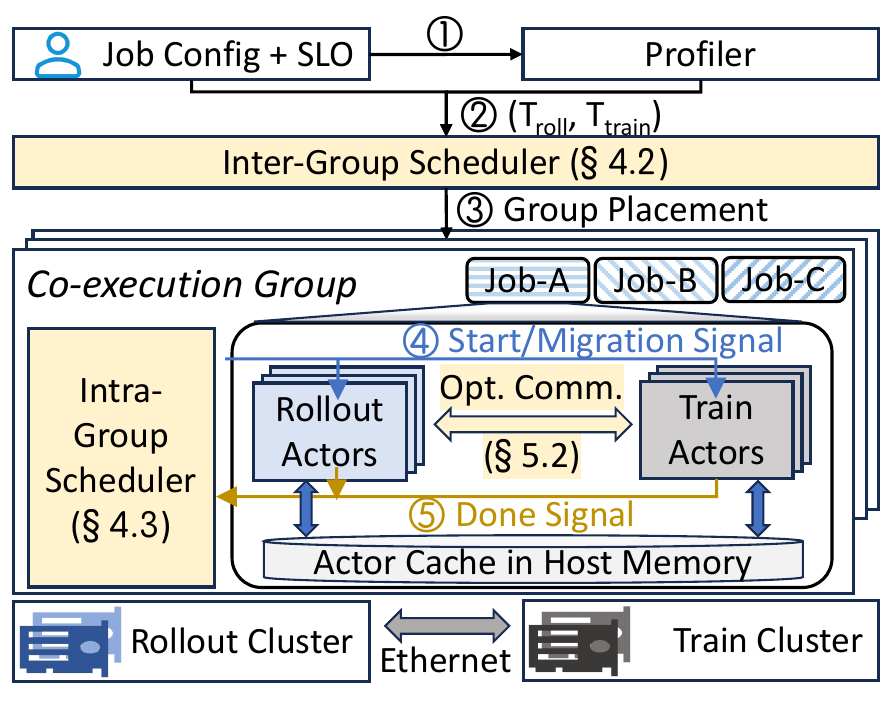}
    \caption{\SystemName system architecture.}
    \label{fig:sys_overview}
\end{figure}

\PHM{Workflow.} 
The system operation follows the closed loop illustrated in \autoref
{fig:sys_overview}. Upon job submission, \SystemName{} first launches a
lightweight profiler (\textcircled{1}) to generate worst-case duration
estimates for the job's rollout and training phases (\textcircled{2}). These
estimates are fed into the inter-group scheduler (\textcircled{3}), which
identifies the optimal co-execution group and resource placement to minimize
marginal cost (\S\ref{subsec:inter-group}). Once placed, the job
comes under the control of the intra-group scheduler (\S\ref
{subsec:intra-group}). This runtime controller orchestrates the round-robin
meta-iteration (\textcircled{4}), enforcing the phase-centric state
management and triggering long-tail 
migrations based on real-time feedback from the runtime hooks (\S\ref{subsec:phase_centric_control}).
Once a phase completes, \SystemName offloads its state to 
the actor cache in host DRAM, releases GPU resources, and subsequently launches 
the next job's waiting phase from the scheduler queue (\textcircled{5}).


\PHM{Isolation and Fault Tolerance.} 
To ensure production-grade reliability, \SystemName{} enforces strict fault
isolation. Each job owns a dedicated Ray~\cite
{moritz2018ray} instance with its isolated runtime environment.
Jobs communicate exclusively with the scheduler via the Redis~\cite
{redis} channel and never directly with one another. Consequently, a crash in one job is fully contained within its pod, preventing
error propagation and ensuring the stability of other jobs within
the same co-execution group.


\section{Evaluation}
\label{sec:eval}

We evaluate \SystemName to answer the following key research
questions. \textbf{RQ1 (Co-Execution Efficacy):} How effectively
does \SystemName's co-execution mechanism manage job groups with diverse
phase profiles (\S\ref{eval:microbenchmark})? \textbf{RQ2 (Performance
Breakdown):} What are the individual contributions of \SystemName's key
optimizations (i.e., long-tail migration, topology-aware model sync) to
overall system performance (\S\ref{eval:ablation})? \textbf{RQ3 (Performance
at Scale):} How does \SystemName perform under a real-world
production workload trace, in terms of cluster-level utilization and
provisioning cost (\S\ref{eval:end2end})? \textbf{RQ4 (Scheduling Quality):}
How efficient and how close to optimal is \SystemName's scheduler (\S\ref
{eval:global_sched})?

\subsection{Experimental Setup}
\label{subsec:experimental_setup}

\PHB{Cluster Setup.} 
Our experimental testbed consists of two geo-distributed, heterogeneous
clusters, where the training cluster (\texttt{Cluster-T}) is equipped with
compute-optimized NVIDIA H800 GPUs, while the rollout cluster (\texttt
{Cluster-R}) is composed of cost-effective H20 GPUs. 
The internal fabric of each cluster is a high-speed 400~Gbps InfiniBand network.
However, the two clusters are connected via a bandwidth-constrained 20~Gbps 
Ethernet link. \autoref{tab:gpu_price} details the hardware specifications 
and hourly costs, where an H800 GPU is 2.85$\times$ more expensive than
an H20 GPU.

\PHM{Workloads.} 
We construct our workloads based on real-world traces collected from a production
cluster. For \textbf{micro-benchmarks} (\S\ref{eval:microbenchmark}--\S\ref
{eval:ablation}), we define a suite of five representative job types (\autoref
{tab:job_config}) using Qwen~\cite{yang2025qwen3technicalreport} models
(7B--32B) with varying batch sizes, sequence lengths, and 
rollout/training GPUs, covering both single-turn RLVR (on DeepMath-103K~\cite
{shao2024deepseekmath} dataset) and multi-turn agentic reasoning
(on Math-Orz57K~\cite{hu2025open} dataset). For \textbf{at-scale evaluation}
(\S\ref{eval:end2end}), we replay a two-week trace comprising 200 heterogeneous jobs. 
This trace features high variance in model sizes (3B--32B) and diverse datasets spanning
mathematics~\cite{shao2024deepseekmath}, software engineering~\cite
{pan2024trainingsoftwareengineeringagents}, games~\cite{frozen_lake}, and
other in-house datasets.

\PHM{Baselines.} 
We compare \SystemName against three baselines.
\begin{itemize}[topsep=3pt, leftmargin=1.5em, noitemsep, nolistsep, parsep=2pt, partopsep=0pt]
  \item \underline{\textit{Solo Disaggregation (Solo-D):}} The standard disaggregation practice
         where jobs are executed on dedicated rollout and training pools without time-multiplexing.
  \item \underline{\textit{Co-location (veRL~\cite{sheng2025hybridflow}):}} The traditional 
        monolithic approach, where all phases execute on the high-performance training 
        cluster (\texttt{Cluster-T}) using the popular veRL framework. It avoids network 
        bottlenecks but suffers from hardware resource mismatch.
  \item \underline{\textit{Gavel+~\cite{narayanan2020heterogeneity}:}} 
        An enhanced version of the heterogeneity-aware Gavel scheduler~\cite{narayanan2020heterogeneity}, modified to support RL post-training. Gavel+ optimizes resource allocation at the job level (calculating optimal GPU fractions) but lacks fine-grained control to 
        interleave phase-level executions.
\end{itemize}

\begin{table}[t]\footnotesize
    \centering
    \begin{tabular}{ccccccc}
        \toprule
        \textbf{Job} & \textbf{Turns} & \textbf{Model} & \textbf{Len\footnotemark} & \textbf{Bsz} & $\bf N_T$ & $\bf N_R$\\
        \midrule
        Type-A & Single-Turn & Qwen-2.5-7B & 8K & 256 & 8 & 8\\
        Type-B & Single-Turn & Qwen-2.5-14B & 8K & 256 & 8 & 8\\
        Type-C & Single-Turn & Qwen-2.5-32B & 8K & 256 & 16 & 16\\
        Type-D & Multi-Turn & Qwen-3-8B & 8K$^*$ & 256 & 8 & 8\\
        Type-E & Multi-Turn & Qwen-3-14B & 16K$^*$ & 64 & 8 & 8\\
        \bottomrule
    \end{tabular}
    \caption{Job configurations in experiments, $N_T, N_R$ are the corresponding numbers of training/rollout GPUs.}
    \label{tab:job_config}
\end{table}
\footnotetext{For multi-turn workloads, Len refers to a per-turn output length.}

\subsection{Micro-Benchmarks}
\label{eval:microbenchmark}


\begin{figure*}[tb]
    \centering
    \begin{minipage}[t]{0.32\textwidth}
        \centering
        \includegraphics[width=\linewidth]{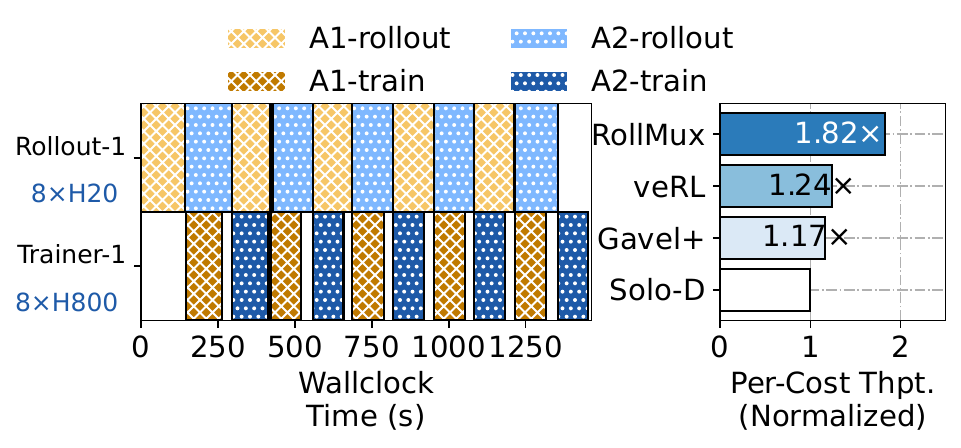}
        \subcaption{Temporal Mux (single-turn, Type-A$\times 2$).}
        \label{fig:microbench_1}
    \end{minipage}
    \hfill
    \begin{minipage}[t]{0.32\textwidth}
        \centering
        \includegraphics[width=\linewidth]{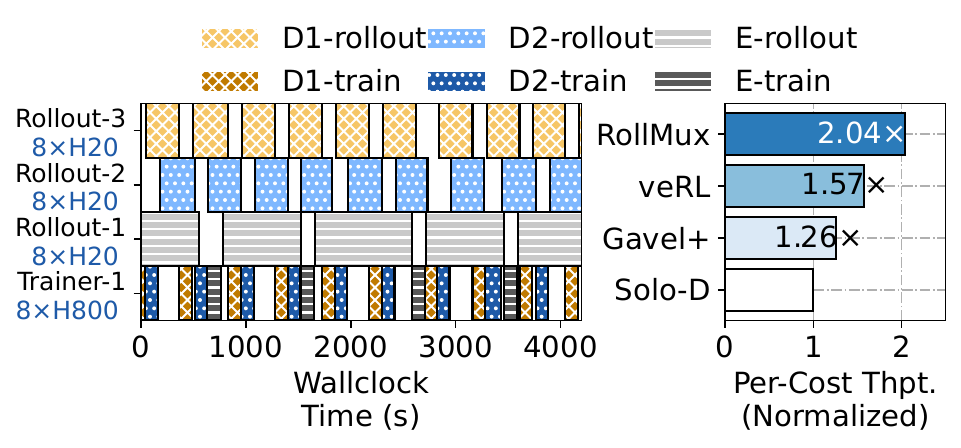}
        \subcaption{Train Mux (multi-turn, Type-D$\times 2$ + E).}
        \label{fig:microbench_2}
    \end{minipage}
    \hfill
    \begin{minipage}[t]{0.32\textwidth}
        \centering
        \includegraphics[width=\linewidth]{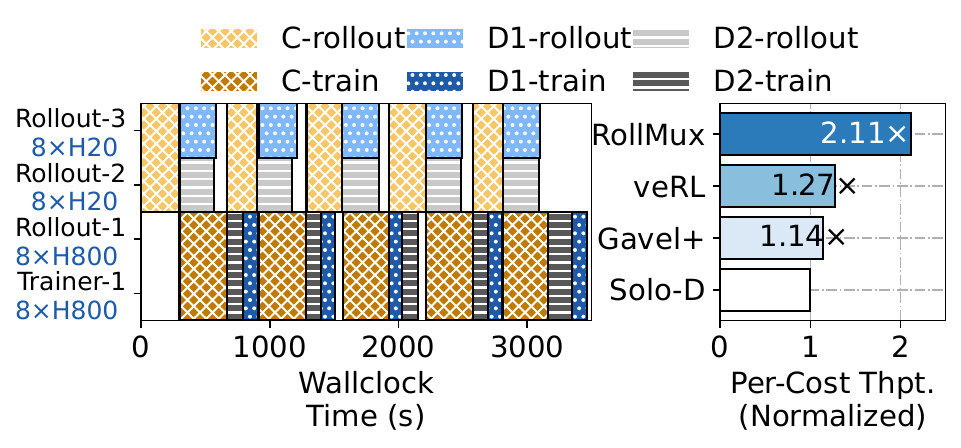}
        \subcaption{Spatial Mux (mixed, Type-C + D$\times2$).}
        \label{fig:microbench_3}
    \end{minipage}
    \caption{Micro-benchmarking results. For each benchmark, the left panel is a gantt chart showing the co-execution timeline; the right panel quantifies the benefit, in which \SystemName achieves $1.82-2.11\times$ higher cost-efficiency. }
    \vspace{-0.5cm}
    \label{fig:microbenchmarks}
\end{figure*}

To demonstrate how \SystemName{} effectively reclaims dependency bubbles (\S\ref{subsec:intra-group}), we conduct three micro-benchmarks in different multiplexing 
scenarios and measure cost efficiency (throughput per
dollar) against the baselines.

\PHM{Temporal Multiplexing.} 
First, we evaluate co-executing two jobs with similar structures (Type-A) via temporal multiplexing, representing an ideal case where jobs are fully complementary. As shown
in \autoref{fig:microbench_1}, \SystemName perfectly interleaves their
execution, keeping both the rollout and training clusters fully utilized.
Consequently, \SystemName improves cost-efficiency by $82\%$, $55.6\%$, and $46.8\%$
over Solo-D, Gavel+, and veRL, respectively. The baselines fall short because 
Solo-D and Gavel+ leave one resource pool idle at all times, while the monolithic veRL underutilizes its expensive H800 compute power during memory-bound rollout phases.

\PHM{Handling Rollout-Heavy Jobs.} 
Next, we target rollout-heavy workloads by co-scheduling two Type-D jobs ($T^
{\text{roll}}_D\approx 2.5 T^{\text{train}}_D$) and one Type-E job ($T_E^
{\text{roll}}\approx 6 T_E^{\text{train}}$). 
In this scenario, \SystemName scales the rollout pool to 24 H20 GPUs, dedicating an inference node to each job's rollout phase while time-multiplexing a single H800 training node for all training phases in a round-robin sequence (\autoref{fig:microbench_2}). This schedule achieves $104\%$, 
$61.9\%$, and $29.9\%$ higher cost-efficiency than Solo-D, Gavel+, and veRL. Solo-D and Gavel+ perform poorly as the prolonged rollout phases force the expensive training nodes to sit idle for extended periods. 


\PHM{Spatial Multiplexing.} Finally, we evaluate \SystemName's ability to handle heterogeneity
by co-scheduling one large Type-C job (requiring $16\times \text{H20}+16\times\text{H800}$) and two smaller Type-D jobs (each requiring $8 \times \text{H20}+ 16 \times \text{H800}$). As depicted in \autoref{fig:microbench_3}, \SystemName{} identifies the idle resources created by the large job's rollout phase and strategically ``packs'' the two smaller jobs into these bubbles. This dynamic spatial packing maximizes aggregate utilization, delivering $111\%$, $85.1\%$, and $66.1\%$ higher cost-efficiency compared to Solo-D, Gavel+, and veRL. 
This result highlights a critical advantage: unlike the baselines, \SystemName{} 
can dynamically consolidate diverse jobs to available capacity.


\PHM{Interference Overhead.} 
To quantify the cost of co-execution, we measure the throughput degradation
caused by inter-job contention (\autoref{tab:speed_overhead}). Because the
inter-group scheduler proactively prunes placements that would violate
residency or performance constraints, \SystemName{} incurs a minimal 5–-9\% overhead
compared to solo execution. Even compared to an idealized co-location upper bound where
each job runs all phases exclusively on expensive H800 GPUs with zero network cost, 
the throughput gap remains a modest
9.0–20.0\%, confirming that the heavy lifting of context switching and synchronization is effectively masked by our optimizations, and bad placements potentially causing contention are precluded.
\vspace{-0.2cm}




\begin{table}[t]\footnotesize
    \centering
    \begin{tabular}{cccc}
        \toprule
        \textbf{Micro-benchmark} & \textbf{Solo Disaggregation} & \textbf{Ideal} & \textbf{\SystemName} \\
        \midrule
        (a) Temporal Mux & 1.00 & 1.07 & 0.98\\
        (b) Train Mux & 1.00 & 1.07 & 0.95\\
        (c) Spatial Mux & 1.00 & 1.11 & 0.91\\
        \bottomrule
    \end{tabular}
    \caption{Normalized training throughput, showing \SystemName incurs less than a 10\% overhead compared to isolated execution (baseline 1.0). For reference, `Ideal' represents the performance ceiling from co-locating all phases on H800.}
    \label{tab:speed_overhead}
\end{table}

\subsection{Ablation Study}
\label{eval:ablation}

We next break down \SystemName{}'s performance to quantify the individual contributions of its key runtime optimizations.

\begin{figure}[tb]
    \centering
    \includegraphics[width=\linewidth]{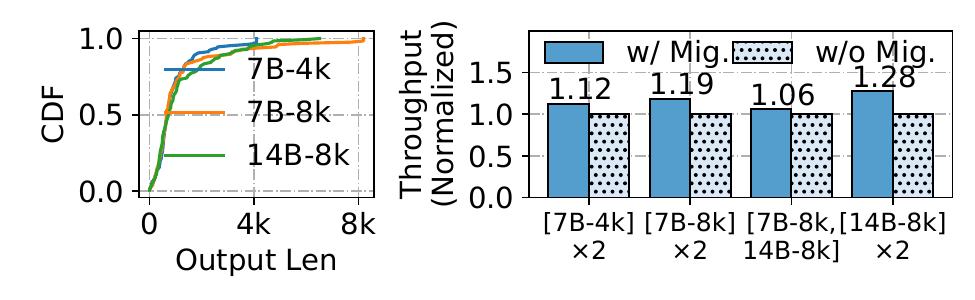}
    \caption{\textbf{Left}: the long-tail distribution of LLM generation length in the rollout phase. \textbf{Right}: effectiveness of request migration in mitigating long-tail rollouts.}
    \label{fig:eval_migration}
\end{figure}

\PHM{Long-Tail Migration.} First, we evaluate the effectiveness of request migration in 
neutralizing the stochasticity of RL rollout (\S\ref{subsec:intra-group}). As illustrated in \autoref{fig:eval_migration}-left, the generation length of rollout requests exhibits a pronounced heavy-tailed distribution across all model sizes and output lengths, 
where a small fraction of "straggler" requests persist long after the majority have finished. \autoref{fig:eval_migration}-right demonstrates that enabling request
 migration effectively reclaims the capacity of ``skewness bubbles.'' By
 preemptively migrating the tail-bound requests to a small subset of
 GPUs, \SystemName{} allows the next job’s rollout phase to begin immediately on the freed majority of resources, improving the end-to-end throughput by 1.06$\times$ to 1.28$\times$. Notably, the gains are
 most pronounced for workloads with longer output sequences (e.g., 14B-8k)
 where the straggler effect is amplified. The benefit is slightly more modest
 when pairing jobs with highly dissimilar characteristics (e.g., 7B-8k and
 14B-8k), as the natural variance in their phase durations already mitigates
 some contention.




\begin{figure}[tb]
    \centering
    \includegraphics[width=0.95\linewidth]{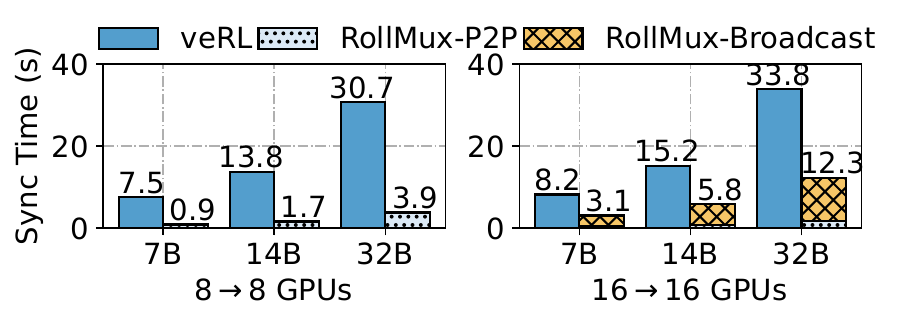}
    \caption{Model synchronization time. \textbf{Left}: single-node, from 8 H800 to 8 H20 GPUs. \textbf{Right}: multi-node, from 16 H800 to 16 H20 GPUs. \SystemName's topology-aware model sync is up to $8.33\times$ faster compared to veRL~\cite{sheng2025hybridflow}. }
    \label{fig:comm_opt}
\end{figure}

\PHM{Topology-Aware Model Sync.}
Next, we evaluate the efficiency of \SystemName's topology-aware model
synchronization scheme in geo-disaggregated setups (\S\ref
{subsec:commopt}). As shown in \autoref{fig:comm_opt}, for a single-node
update (8 H800s $\rightarrow$ 8 H20s), \SystemName achieves 7.87$\times$--8.33$\times$ speedup over
veRL. This significant improvement stems from its hierarchical strategy: \SystemName{}
transmits exactly one copy of the model parameters across the slow inter-cluster 
link and leverages the high-bandwidth local NVLink 
fabric for the final intra-cluster broadcast. In contrast, the baseline is 
bottlenecked by redundantly fetching independent copies for every rollout GPU.
This advantage scales robustly: even in a multi-node setting (16 H800s
$\rightarrow$ 16 H20s), \SystemName{} maintains 2.62$\times$--2.75$\times$ speedup 
(\autoref{fig:comm_opt}),
confirming that our topology-aware protocol effectively masks the bandwidth
limitations inherent to disaggregated clusters.



\begin{figure*}[t]
    \centering
    \begin{minipage}[t]{0.32\textwidth}
        \centering
        \includegraphics[width=\linewidth, trim=0 10 0 30, clip]{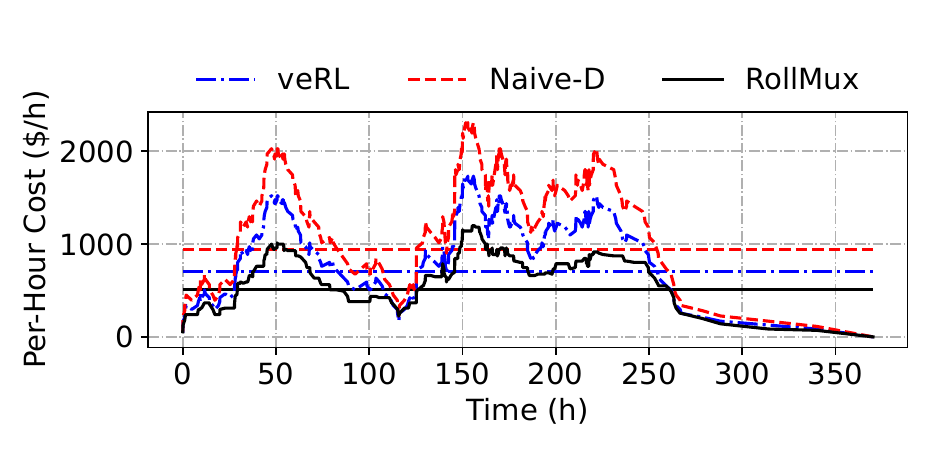}
        \subcaption{Cluster provisioning cost.}
        \label{fig:wild_time_cost}
    \end{minipage}
    \hfill
    \begin{minipage}[t]{0.32\textwidth}
        \centering
        \includegraphics[width=\linewidth, trim=0 10 0 30, clip]{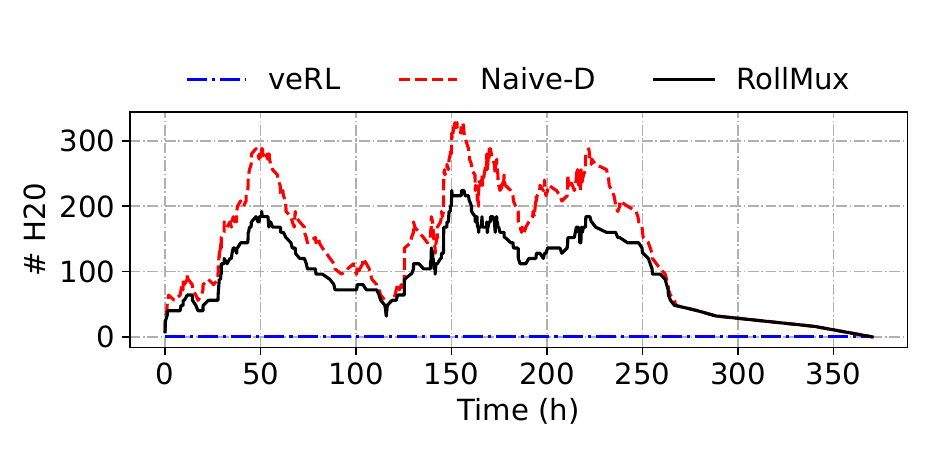}
        \subcaption{Number of rollout (H20) GPUs.}
        \label{fig:wild_time_h20}
    \end{minipage}
    \hfill
    \begin{minipage}[t]{0.32\textwidth}
        \centering
        \includegraphics[width=\linewidth, trim=0 10 0 30, clip]{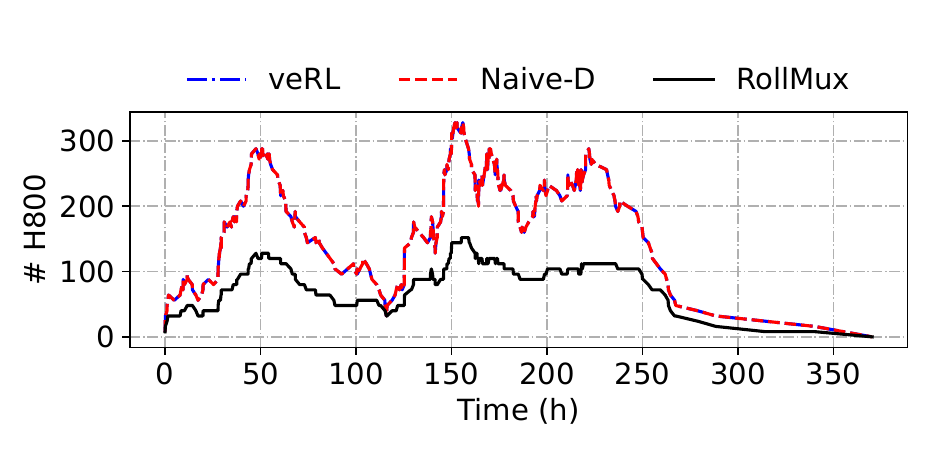}
        \subcaption{Number of training (H800) GPUs.}
        \label{fig:wild_time_h800}
    \end{minipage}
    \caption{\textbf{[Testbed]} Cluster provisioning cost and GPU usage of \SystemName and baselines under real-world production workloads, where \SystemName reduces total provisioning cost by $1.38\times$ and $1.84\times$ compared to veRL and Solo-D, respectively.}
    \label{fig:end2end_wild}
\end{figure*}
\begin{figure*}[t]
    \centering
    \begin{minipage}[t]{0.32\textwidth}
        \centering
        \includegraphics[width=\linewidth, trim=0 10 0 30, clip]{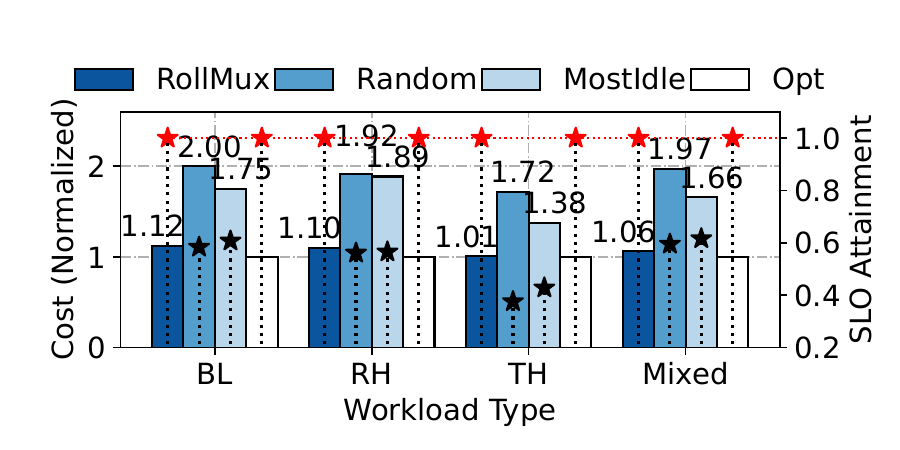}
        \subcaption{Sensitivity to workload types.}
        \label{fig:ablation_type}
    \end{minipage}
    \hfill
    \begin{minipage}[t]{0.32\textwidth}
        \centering
        \includegraphics[width=\linewidth, trim=0 10 0 30, clip]{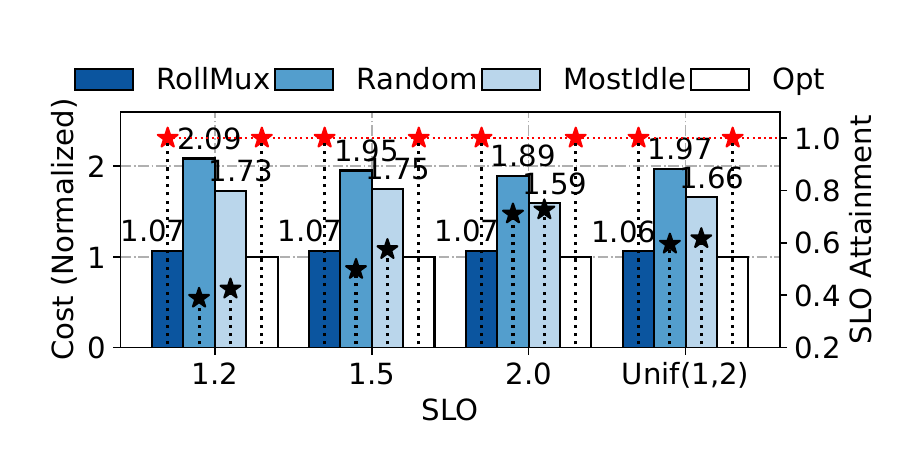}
        \subcaption{Sensitivity to job SLOs.}
        \label{fig:ablation_slo}
    \end{minipage}
    \hfill
    \begin{minipage}[t]{0.32\textwidth}
        \centering
        \includegraphics[width=\linewidth, trim=0 10 0 30, clip]{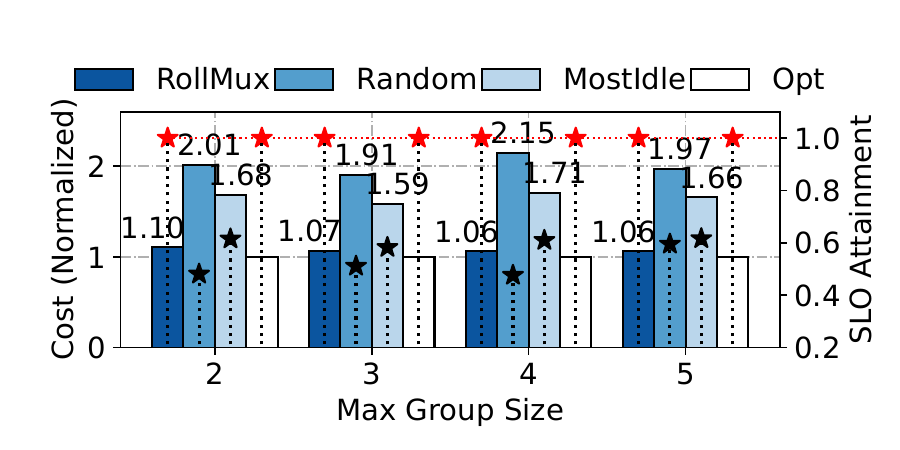}
        \subcaption{Sensitivity to group residency.}
        \label{fig:ablation_grp_size}
    \end{minipage}
    \caption{\textbf{[Simulation]} Sensitivity analysis of \SystemName's inter-group scheduler.}
    \vspace{-0.5cm}
    \label{fig:scheduler_ablation}
\end{figure*}

\subsection{\SystemName at Scale}
\label{eval:end2end}
To evaluate \SystemName's performance at production-scale, we replay a two-week trace from one of our cluster tenants. The trace comprises 200 highly heterogeneous RL post-training jobs using Qwen-family models, including both single- and multi-turn interaction patterns on diverse datasets. Among them, the model sizes range from 3B to 32B, the maximum response lengths range from 4k to 32k tokens (mean: 12.1k), and the mean job duration is 27.9 hours.
We assign each job an SLO sampled uniformly from $(1, 2)$ relative to its solo runtime. We compare \SystemName against the industry-standard solo disaggregation (Solo-D), with 1:1 rollout and training GPUs, and the monolithic co-located baseline (veRL).

In terms of cluster provisioning cost, \SystemName spends only \$510 per hour to accommodate all jobs, a 1.84$\times$ and 1.38$\times$ reduction compared to Solo-D and veRL, respectively, while meeting all job SLOs (\autoref{fig:wild_time_cost}). This cost efficiency directly stems from our scheduling algorithm (\autoref{algo:inter_scheduling}) that minimizes the marginal cost upon job arrivals.

In terms of resource efficiency, \SystemName reduces dependency bubbles by 24.4\% on the rollout cluster and 43.1\% on the training cluster compared to solo disaggregation. The improvement is more pronounced for training because the workloads are typically rollout-heavy, leaving more bubbles on the training GPUs (\autoref{fig:wild_time_h20} and \autoref{fig:wild_time_h800}). By tightly packing jobs, \SystemName requires a peak of only 152 H800 GPUs for training, a 2.16$\times$ reduction from the 328 H800s needed by both veRL and solo disaggregation. For rollout, \SystemName's peak usage is 216 H20 GPUs, a 1.52$\times$ reduction compared to the 328 H20s required by solo disaggregation. Although the co-located veRL baseline uses no separate rollout GPUs, \SystemName is still 1.38$\times$ more cost-effective overall. It achieves this by offloading memory-bandwidth-intensive rollout phases to cheaper H20 GPUs via disaggregation while filling the resulting dependency bubbles by co-scheduling.
\vspace{-0.3cm}

\subsection{Scheduler Performance}
\label{eval:global_sched}

We finally evaluate the optimality and scalability of \SystemName's inter-group scheduler (\S\ref{subsec:inter-group}) via large-scale trace simulation.

\PHB{Experimental Setup.}
We use job arrival patterns from a 300-job, 580-hour segment of the Microsoft Philly multi-tenant training cluster trace~\cite{jeon2019analysis}, the average job duration is 14.4 hours, and the longest is 142.9 hours. While the trace dictates arrival times and durations, we synthesize the job characteristics to model modern RL post-training workloads. As detailed in \autoref{tab:scheduler_jobs}, we define three job profiles based on the ratio of rollout time ($T_{\text{roll}}$) to training time ($T_{\text{train}}$). (1) \textbf{Balanced (BL):} Balanced $T_{\text{roll}}$ and $T_{\text{train}}$, representative of single-turn workloads like RLHF~\cite{ouyang2022training} or RLVR~\cite{shao2024deepseekmath}; (2) \textbf{Rollout-Heavy (RH):} $T_{\text{roll}} \gg T_{\text{train}}$, modeling multi-turn workloads such as agentic reasoning~\cite{pan2024trainingsoftwareengineeringagents}; and (3) \textbf{Train-Heavy (TH):} $T_{\text{train}} \gg T_{\text{roll}}$ for evaluation completeness, but is rare in real-world RL training.

For each profile, we generate jobs of three sizes (Small, Medium, Large), resulting in nine distinct job configurations. We test each profile individually and also use a \textbf{Mixed} workload, which contains a uniform mix of all nine configurations, to simulate a realistic production environment.

\PHM{Baselines.}
We compare \SystemName against following baselines:

\begin{itemize}[topsep=3pt, leftmargin=1.5em, noitemsep, nolistsep, parsep=2pt, partopsep=0pt]
    \item \underline{\textit{Offline Optimal (Opt).}} Assigns arriving jobs to \textit{offline optimal} placements found via a brute-force search over all possible groupings and placements--including job re-ordering and re-grouping that are infeasible in online deployments--and serves as a theoretical upper bound.
    \item \underline{\textit{Random.}} Assigns arriving jobs to a random group (or a new one) that can accommodate it. The job is then placed on random rollout and train nodes within this group.
    \item \underline{\textit{Greedy (Most-Idle).}} Assigns arriving jobs to the group with the highest idle-time percentage. Within that group, it places the job on the most idle rollout and train nodes.
\end{itemize}

\PHM{Sensitivity Analysis.} To determine how sensitive the scheduler's quality is to its key parameters, we vary a single parameter while all others are set to a default configuration\footnote{Default configuration: mixed workload types, heterogeneous SLOs drawn from \texttt{Unif(1,2)}, and a max group residency of 5.}.

\textbf{\textit{Impact of Workload Characteristics.}}
\autoref{fig:ablation_type} shows that \SystemName's near-optimal performance holds across all workload types. It consistently achieves 100\% SLO attainment with a cost overhead of just 1.01$\times$–1.12$\times$ relative to the optimal. In contrast, the baselines perform poorly across different workloads, with Greedy slightly outperforms the Random strategy. The Random strategy's cost is 1.72$\times$–2.00$\times$ optimal with only 37–58\% SLO attainment, while the Greedy scheduler is slightly better but still inadequate, reaching 1.38$\times$–1.89$\times$ optimal cost for 42–61\% SLO attainment.

\textbf{\textit{Impact of Job SLOs.}}
We vary the job SLO requirements, testing both uniform SLOs (all jobs set to 1.2, 1.5, or 2.0) and heterogeneous, job-specific SLOs drawn from \texttt{Unif(1, 2)}. As shown in \autoref{fig:ablation_slo}, \SystemName's performance is highly stable against SLO tightness, always achieving 100\% attainment with a consistent, near-optimal cost. Conversely, the baselines are highly sensitive to the SLO target and more expensive (1.59$\times$–2.09$\times$ optimal). As the SLO target loosens from 1.2 to 2.0, the SLO attainment for Random and Greedy improves from 38\%/43\% to 71\%/73\%, respectively. This shows that while looser SLOs make it easier for naive heuristics to succeed by chance, they still fail to provide guarantees.

\textbf{\textit{Impact of Group Residency.}}
Since node memory capacity directly limits the group size,  we evaluate its impact by varying the maximum allowed group size from 2 to 5. \autoref{fig:ablation_grp_size} shows that performance is relatively insensitive to the maximum group size for all methods. Across all configurations, \SystemName consistently maintains the lowest cost and 100\% SLO attainment, while the baselines remain significantly defected (only 48\%–61\% SLO attainment) and more expensive (1.59$\times$–2.15$\times$ optimal). This suggests that even small group sizes (e.g., 2 or 3) provide sufficient packing flexibility for \SystemName to find efficient placements; larger groups do not bring improved performance or cost efficiency.

\PHM{Scheduler Scalability and Latency.} We evaluate \SystemName's decision latency and scalability, with results presented in \autoref{tab:sched_overhead}. \SystemName demonstrates near-linear scalability: its decision time is only 591 ms for 2,000 jobs, confirming that \autoref{algo:inter_scheduling} efficiently handles production-scale workloads. In stark contrast, the brute-force optimal solver exhibits exponential growth in latency, exceeding five hours for just 13 jobs--rendering it infeasible for any practical workload size.
\vspace{-0.3cm}

\begin{table}[t]\footnotesize
  \centering
  \begin{tabular}{l ccccccc}
    \toprule
    \textbf{Decision} & \multicolumn{6}{c}{\textbf{Number of Concurrent Jobs}} \\
    \cmidrule(l){2-8}
    \textbf{Lat. (ms)} & \textbf{5} & \textbf{9} & \textbf{13} & \textbf{100} & \textbf{500} & \textbf{1000} & \textbf{2000} \\
    \midrule
    \textbf{\SystemName} & 5.6 & 6.5 & 7.6 & 41.9 & 198 & 318 & 591 \\
    \textbf{Opt.} & 113 & >1min\textsuperscript{\textdagger} & \textgreater 5h\textsuperscript{\textdagger} & ---\textsuperscript{\textasteriskcentered} & ---\textsuperscript{\textasteriskcentered} & ---\textsuperscript{\textasteriskcentered} & ---\textsuperscript{\textasteriskcentered} \\
    \bottomrule
    \multicolumn{8}{l}{\footnotesize \textsuperscript{\textdagger} Represents latency exceeding 1 minute and 5 hours, respectively.} \\
    \multicolumn{8}{l}{\footnotesize \textsuperscript{\textasteriskcentered} Not applicable; computation is intractable at this scale.} \\
  \end{tabular}
  \caption{Decision latency (ms) vs. number of concurrent jobs. \SystemName scales well; Brute-force Opt's latency grows exponentially and quickly becomes impractical.}
  \label{tab:sched_overhead}
\end{table}

\section{Related Work}
\label{sec:related_work}

\PHB{Deep Learning Schedulers.}
Extensive research has focused on scheduling for general-purpose deep learning clusters. These works aim to improve fairness~\cite{xiao2018gandiva, mahajan2020themis, gu2019tiresias}, GPU sharing efficiency~\cite{xiao2020antman, weng2023beware}, heterogeneity awareness~\cite{narayanan2020heterogeneity, shen2025xsched}, and job goodput~\cite{qiao2021pollux, zheng2023shockwave}. However, these systems universally treat the entire job as the atomic unit of scheduling and assume stable, predictable iteration times. \SystemName is the first phase-centric co-scheduling system for stochastic RL post-training jobs.

\PHM{RL Post-Training Frameworks.}
Recent systems have focused on optimizing the performance of a \textit{single, individual} RL post-training job. While early frameworks used static resource partitioning~\cite{hu2024openrlhf, yao2023deepspeed, kuchaiev2019nemo}, subsequent work improved utilization via co-location~\cite{sheng2025hybridflow}, multi-controller designs~\cite{wang2025distflow}, long-tail mitigation~\cite{zhong2024optimizing,gao2025rollpacker}, and asynchronous algorithms~\cite{fu2025areal,zhong2025streamrl,han2025asyncflow}. Different from these single-job optimizations, \SystemName addresses the complementary part by taking a global, cluster-level perspective, orchestrating multiple concurrent jobs via co-scheduling.

\PHM{Disaggregated Systems.} \SystemName builds on the principle of disaggregation, a concept explored in OS kernels~\cite{shan2018legoos} and serving systems~\cite{hu2024inference, zhong2024distserve, patel2024splitwise}. The most direct parallel is the prefill-decode disaggregation in LLM inference~\cite{hu2024inference, zhong2024distserve, patel2024splitwise}. Within this domain, recent serving systems have also explored efficient resource management, but all in a single-job scope~\cite{feng2025windserve, cui2025optimizingsloorientedllmserving,zhu2025nanoflowoptimallargelanguage, hong2025semi}. In contrast, \SystemName introduces the first scheduling framework specifically designed for multi-tenant clusters, a scenario not addressed by prior works.

\vspace{-0.3cm}

\section{Conclusion}
\label{sec:conclusion}

This paper presents \SystemName, a multi-tenant cluster scheduler tailored for rollout-training disaggregated RL post-training. \SystemName introduces a two-tier scheduling mechanism that near-optimally partitions RL jobs into co-execution groups and orchestrates their execution in a tightly-woven pattern. This approach effectively reduces dependency bubbles caused by disaggregation, thereby minimizing cluster provisioning costs. Extensive evaluation using real-world production traces on disaggregated clusters with up to 328 GPUs each demonstrates that \SystemName achieves up to $1.84\times$ cost savings while maintaining 100\% performance SLO attainment.



\bibliographystyle{plain}
\bibliography{sample}

\clearpage
\section*{Appendix}
\PHB{Proof of Meta-iteration Schedule's Optimality.}\label{sec:appendix_proof}
To show the utilization-optimality of the meta-iteration schedule for non-overloaded groups, we first formally define the utilization of a group with meta-iteration time $T_{\text{meta}}$, where its rollout ($U_R$) and training utilization ($U_T$) are:
$$\textstyle
U_R = \frac{\sum_{j\in  J_{G}}T^{\text{roll}}_{j}}{T_{\text{meta}}},\quad \text{and} \quad U_T = \frac{\sum_{j\in  J_{G}}T^{\text{train}}_{j}}{T_{\text{meta}}}.
$$

Since the meta-iteration schedule repeats, it suffices to analyze a single meta-iteration where every job’s rollout and training phase executes once. Recall that $T_{G}^{\text{cycle}} = \max_{j \in  J_{G}} T^{\text{solo}}_{j}$ is the solo iteration time of the slowest job, and let $j_1$ be the job attaining this maximum so that $T_{G}^{\text{cycle}} = T^{\text{roll}}_{j_1} + T^{\text{train}}_{j_1}$. For simplicity, the following proof considers only one rollout node and one training node, while the optimality generalizes to multi-node settings. By the definition of a non-overloaded group, we have
$$\textstyle T_{G}^{\text{cycle}} = T^{\text{roll}}_{j_1} + T^{\text{train}}_{j_1} \ge \max\left(\sum_{j\in J_{G}}T^{\text{roll}}_{j}, \sum_{j\in J_{G}} T^{\text{train}}_{j}\right). $$

This directly implies that the total rollout work of all other jobs can fit within $j_1$'s training phase, and vice-versa:
$$\textstyle
\sum_{j\in J_{G}\setminus j_1}T^{\text{roll}}_{j} \le T^{\text{train}}_{j_1}, \quad \sum_{j\in J_{G}\setminus j_1}T^{\text{train}}_{j} \le T^{\text{roll}}_{j_1}.
$$

Consequently, job $j_1$'s execution time decides the entire schedule's cycle time, yielding the cycle time as $T_{j_1}^{\text{solo}}$.

Conversely, any schedule that repeats a job is sub-optimal for a non-overloaded group because it extends the meta-iteration's duration but adds proportionally less work, leading to a net decrease in resource utilization. We show this by trying to add a repetition of any job $k$ to the meta-iteration schedule. Repeating job $k$ adds work to both pools and necessarily prolongs the cycle time by at least $T^{\text{solo}}_{k}$ due to waiting for the longest $j_1$, causing the new utilization $U'$ lower than the original $U$. We can bound the change in utilization, $\Delta U = U' - U$, as follows. The change for the rollout and the training pool, $\Delta U_R$ and $\Delta U_T$, is bounded by:
{
\small
\begin{align}\nonumber
    \textstyle
    & \Delta U_R = U_R' - U_R \le \frac{{\sum_{j\in{ J_{G}}}T^{\text{roll}}_{j}} + T^{\text{roll}}_{k}}{ T_{j_1}^{\text{solo}} + T^{\text{roll}}_{k}} - \frac{{\sum_{j\in{ J_{G}}}T^{\text{roll}}_{j}}}{T_{j_1}^{\text{solo}}},\\
    \textstyle\nonumber
    & \Delta U_T = U_T' - U_T \le \frac{{\sum_{j\in{ J_{G}}}T^{\text{train}}_{j}} + T^{\text{train}}_{k}}{ T_{j_1}^{\text{solo}} + T^{\text{train}}_{k}} - \frac{{\sum_{j\in{ J_{G}}}T^{\text{train}}_{j}}}{T_{j_1}^{\text{solo}}}.
\end{align}
}

Therefore, the total change in utilization is
{
\small
\begin{equation}\textstyle
    \begin{aligned}
         \Delta U &= \Delta U_R + \Delta U_T = U_R' - U_R + U_T' - U_T\\
     \nonumber &\le \frac{T^{\text{train}}_{k}T^{\text{solo}}_{j_1} - T^{\text{solo}}_{k}\sum_{j\in  J_{G}} T^{\text{train}}_{j}+ T^{\text{solo}}_{j_1} T^{\text{roll}}_{k} - T^{\text{solo}}_{k}\sum_{j\in J_{G}} T^{\text{roll}}_{j}}{T^{\text{solo}}_{j_1}(T^{\text{solo}}_{j_1}+T^{\text{solo}}_{k})} \\
     \textstyle\nonumber &= \frac{T^{\text{solo}}_{k}(T^{\text{solo}}_{j_1} -\sum_{j\in J_{G}} T^{\text{solo}}_{j})}{T^{\text{solo}}_{j_1}(T^{\text{solo}}_{j_1}+T^{\text{solo}}_{k})} \le 0.
    \end{aligned}
\end{equation}
}

This proves that any repetition degrades overall utilization, making the round-robin schedule utilization-optimal for non-overloaded groups.

\PHM{Workloads for Simulation.}
\autoref{tab:scheduler_jobs} details the job profiles used in \S~\ref{eval:global_sched}.

\begin{figure}[bt]
    \centering
    \includegraphics[width=0.95\linewidth]{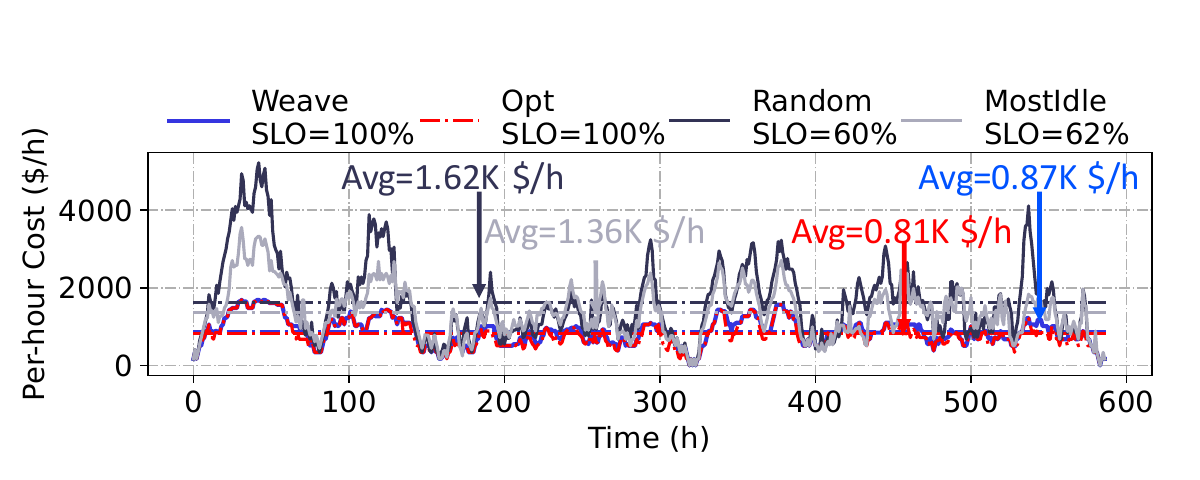}
    \caption{\textbf{[Simulation]} Cost-effectiveness and SLO attainment under a realistic mixed workload. (Workload: mixed, SLO$\sim$\texttt{Unif(1,2)}, max group size=5).}
    \label{fig:time_cost_sim}
\end{figure}

\PHM{[Simulation] End-to-end Performance.} This part reports the end-to-end performance of \SystemName's scheduler under a realistic setting with Mixed workloads with heterogeneous job SLOs. As illustrated in \autoref{fig:time_cost_sim}, \SystemName, achieves 100\% SLO attainment by design, matching the theoretical Offline Optimal scheduler. In contrast, the heuristic-based baselines struggle significantly: the Random and Greedy (Most-Idle) schedulers meet the SLOs for only 60\% and 62\% of jobs, respectively. This is because \SystemName's cost-based optimization (\S~\ref{subsec:intra-group}) explicitly prunes any placements that would violate a job's SLO by assigning it an infinite cost.

In terms of resource efficiency, the average cost of \SystemName is at 0.87K \$/h, only 1.06$\times$ higher than the Offline Optimal. In contrast, the baselines are far more expensive. The Random and Greedy strategies incur average costs of 1.62K and 1.36K \$/h, respectively (1.97$\times$ and 1.66$\times$ of optimal). This high cost is driven by their sub-optimal group partitioning, where their costs spike to over 5K \$/h by scaling out to 1400 GPUs. \SystemName, however, manages load with a peak cost of only $\sim$1.8K \$/h, requiring at most 504 GPUs, demonstrating its effectiveness at identifying SLO-aware, cost-efficient placements in a complex, online setting.

\begin{table}[bt]
\centering
\caption{Job configurations used in the simulation. Each job is defined by its rollout time ($T_{\text{roll}}$) and train time ($T_{\text{train}}$), which are drawn from different uniform distributions.}
\label{tab:scheduler_jobs}
\footnotesize 
\begin{tabular}{@{}llcc@{}} 
\toprule
\textbf{Workload} & \textbf{Size} & \textbf{Rollout Time ($\mathbf{T_{\text{roll}}}$)} & \textbf{Train Time ($\mathbf{T_{\text{train}}}$)} \\
\midrule
\textbf{Balanced}& Small  & \texttt{Unif(50, 100)}   & \texttt{Unif(50, 100)}   \\
\textbf{(BL)} & Medium & \texttt{Unif(100, 200)}  & \texttt{Unif(100, 200)}  \\
& Large  & \texttt{Unif(200, 300)}  & \texttt{Unif(200, 300)}  \\
\midrule
\textbf{Rollout-} & Small  & \texttt{Unif(100, 200)}  & \texttt{Unif(25, 50)}    \\
\textbf{Heavy} & Medium & \texttt{Unif(200, 400)}  & \texttt{Unif(50, 100)}   \\
\textbf{(RH)} & Large  & \texttt{Unif(400, 600)}  & \texttt{Unif(100, 200)}  \\
\midrule
\textbf{Train-} & Small  & \texttt{Unif(25, 50)}    & \texttt{Unif(100, 200)}  \\
\textbf{Heavy} & Medium & \texttt{Unif(50, 100)}   & \texttt{Unif(200, 400)}  \\
\textbf{(TH)} & Large  & \texttt{Unif(100, 200)}  & \texttt{Unif(400, 600)}  \\
\midrule
\textbf{Mixed} & All & All & All\\
\bottomrule
\end{tabular}
\end{table}

\end{document}